\documentclass[twocolumn,showpacs,prb]{revtex4-1}%
\usepackage{amsfonts}
\usepackage{amsmath}
\usepackage{amssymb}
\usepackage{subfigure}
\usepackage{graphicx}
\usepackage{txfonts}%
\providecommand{\U}[1]{\protect\rule{.1in}{.1in}}

\begin{document}

\title{Anomalous caustics and Veselago focusing in 8-\textit{Pmmn} borophene \textit{p-n} junctions with arbitrary junction directions}

\author{Shu-Hui Zhang$^{1}$}
\email{shuhuizhang@mail.buct.edu.cn}
\author{Wen Yang$^{2}$}
\email{wenyang@csrc.ac.cn}

\affiliation{$^{1}$College of Mathematics and Physics, Beijing University of Chemical Technology, Beijing,
100029, China}
\affiliation{$^{2}$Beijing Computational Science Research Center, Beijing 100193, China}

\begin{abstract}

Negative refraction usually demands complex structure engineering while it is very natural for massless Dirac fermions (MDFs) across the
\textit{p-n} junction, this leads to Dirac electron optics. The emergent Dirac materials may exhibit hitherto unidentified phenomenon due to their nontrivial band structures in contrast to the isotropic MDFs in graphene. Here, as a specific example, we explore the negative refraction induced
caustics and Veselago focusing of tilted MDFs across 8-\textit{Pmmn} borophene
\textit{p-n} junctions. To this aim, we develop a technique to effectively
construct the electronic Green's function in \textit{p-n} junctions with
arbitrary junction directions. Based on analytical discussions and numerical calculations, we demonstrate the strong dependence of interference pattern on the junction direction. As the junction direction perpendicular to the tilt
direction, Veselago focusing or normal caustics (similar to that in graphene) appears resting on the doping
configuration of the \textit{p-n} junctions, otherwise anomalous caustics (different from that in graphene)
occurs which is manipulated by the junction direction and the doping
configuration. Finally, the developed Green's function technique is generally promising to uncover the unique transport of emergent MDFs, and the discovered anomalous caustics makes tilted MDFs potential applications in Dirac electron optics.

\end{abstract}
\maketitle

\section{Introduction}

Negative refraction is one unusual class of wave propagation\cite{PendryPRL2000}, which leads to novel interference and hold great potential in new applications\cite{Smith788}. But its realization usually demands complex structure engineering with stringent parameter conditions\cite{Chen2016}. Graphene as the first two-dimensional material, hosts the
relativistic massless Dirac fermions (MDFs) and possesses a lot of exotic
physics and possible applications \cite{CastroRMP2009}. Especially, the two-dimensional nature of graphene is beneficial to the fabrication of the planar \textit{p-n} junction (PNJ), which is the basic component of many electronic devices\cite{,Low2012,C7CS00880E}. The propagation of MDFs across graphene PNJ has a close analogy to optical negative refraction at the surface of metamaterials but exhibits negative refraction in a more simple and tunable manner\cite{CheianovScience2007}. Recently, negative refraction in the graphene PNJ has
been verified experimentally\cite{LeeNatPhys2015,ChenScience2016}. As a result, there is wide interest to study the negative refraction induced interference of MDFs in the PNJ structure\cite{PhysRevLett.100.236801,MoghaddamPRL2010,SilveirinhaPRL2013,ZhaoPRL2013,MilovanovicJAP2015,ncomms15783,zhang2017,PhysRevB.95.214103,PhysRevB.97.205437,PhysRevB.98.205421,PhysRevB.99.094111,PhysRevB.100.041401}.

The great success of graphene attracts people to search for new two-dimensional
materials in which quasiparticles can be described as MDFs \cite{Wehling2014,WangNSR2015}, i.e., Dirac materials. Dirac materials emerge quickly  and usually host very novel MDFs, in which new physics and application potential are expected\cite{Wehling2014,WangNSR2015,S136970211830097X}. The Dirac fermions can be classified finely into four categories\cite{PhysRevX.9.031010}, i.e., type-I, type-I tilted, type-III (critical tilt), and type-II ones. These four categories of Dirac fermions are expected to exhibit distinct physical properties due to different geometries of their Fermi surface. To our knowledge, type-I tilted Dirac fermions has been predicted to appear in very rare systems including quinoid-type graphene and ¦Á-(BEDT-TTF)$_2$I$_3$\cite{PhysRevB.78.045415}, hydrogenated graphene\cite{PhysRevB.94.195423}, and 8-\textit{Pmmn} borophene\cite{PhysRevLett.112.085502,PhysRevB.93.241405,PhysRevB.94.165403,PhysRevB.97.125424}. In such Dirac materials, 8-\textit{Pmmn} borophene as elemental monolayer material exhibits high mobility and anisotropic transport\cite{C7CP03736H}, which has potential applications in electronics and electron optics. The well-established continuum model of 8-\textit{Pmmn} borophene make it be very suitable to the model study. This anisotropy and tilt of type-I Dirac fermions brings about unique features to various physical properties of 8-\textit{Pmmn} borophene, including plasmon
\cite{PhysRevB.96.035410,PhysRevB.98.235430}, the optical conductivity
\cite{PhysRevB.96.155418}, Weiss oscillations \cite{PhysRevB.96.235405}, oblique Klein tunneling\cite{PhysRevB.97.235440}, metal-insulator transition induced by strong electromagnetic radiation\cite{PhysRevB.99.035415}, and RKKY interaction\cite{PhysRevB.99.155418,zhangJMMM2019}. Thus, it is expected that the anisotropy and tilt will strongly affect the negative refraction of MDFs cross 8-\textit{Pmmn} borophene PNJ.

In this study, we investigate the negative refraction induced
interference in 8-\textit{Pmmn} borophene
PNJs. Following the seminal study in graphene\cite{CheianovScience2007}, we focus on two typical interference pattern, i.e, caustics and Veselago focusing. Due to the anisotropy and tilt of MDFs in 8-\textit{Pmmn} borophene, the interference pattern has inevitable dependence on
the junction direction. Hence, we develop a construction technique of the electronic Green's function (GF), which is applicable to the PNJ with an
arbitrary junction direction. We find: (I) Veselago focusing or normal caustics appears resting on the doping
configuration of the PNJ when its junction direction is perpendicular to the tilt
direction of MDFs. (II) To the other junction, anomalous caustics unique to anisotropic and tilted MDFs
occurs, and its pattern strongly depends on the junction direction and the doping
configuration of the PNJ. We expect that the developed GF technique is used to study the other transport properties of continuing emergent MDFs, and the discovered  anomalous caustics can be observed in near future pointing out the potential application of novel MDFs (in 8-\textit{Pmmn} borophene or other Dirac materials) in Dirac electron optics.

The rest of this paper is organized as follows.
In Sec. II, we introduce the model structures and Hamiltonian of the 8-\textit{Pmmn}
borophene PNJ, and present the detailed construction technique of the electronic GF. In Sec. III, according to the junction direction of PNJs, we demonstrate Veselago focusing and anomalous caustics by combing numerical calculations and analytical derivation. Finally, we summarize this study in Sec. IV.

\section{Theoretical formalism}

\subsection{Model and Hamiltonian}

\begin{figure}[ptbh]
\includegraphics[width=1.0\columnwidth,clip]{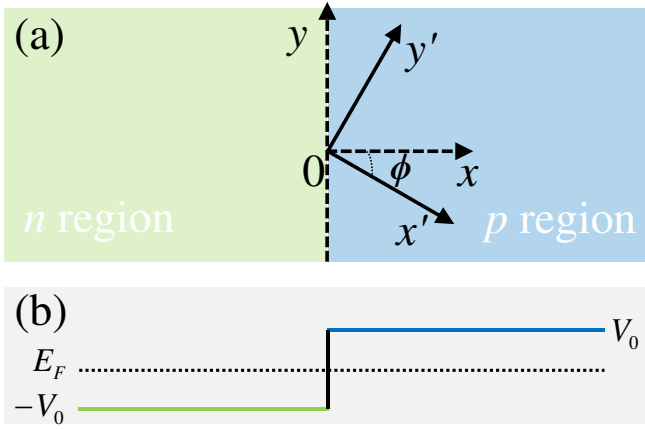}\caption{(a)
Schematic \textit{p-n} junction with the left \textit{n} region and the right
\textit{p} region. Two Cartesian coordinate systems $x-y$ and $x^{\prime
}-y^{\prime}$ are introduced, and $x-y$ has one rotation relative to the
$x^{\prime}-y^{\prime}$ with the angle $\phi$. The normal (tangential)
direction of the junction interface defines the $x$ ($y$) axis of the
coordinate system $x-y$ while the coordinate system $x^{\prime}-y^{\prime}$ is
used to express the intrinsic Hamiltonian. So $\phi$ is used to denote the
junction direction. (b) The energy position $-V_{0}$ ($V_{0}$) for the Dirac
point of the \textit{n} (\textit{p}) region relative to the aligned Fermi
energy $E_{F}$.}%
\label{structure}%
\end{figure}

We introduce the Cartesian coordinate system $x^{\prime}-y^{\prime}$
accompanying the intrinsic Hamiltonian $\hat{H}_{0}$ of 8-\textit{Pmmn}
borophene. In 8-\textit{Pmmn}
borophene, $\hat{H}_{0}$ describes the anisotropic and tilted MDFs, and has the valley dependence. However, we focus the new physics induced by the anisotropy and tilt, so the valley dependence of $\hat{H}_{0}$ is neglected in our model study. Around
the Dirac point, $\hat{H}_{0}$ has the form \cite{PhysRevB.94.165403,PhysRevB.96.035410,PhysRevB.96.235405}%

\begin{equation}
\hat{H}_{0}=v_{1}\sigma_{1}\hat{p}_{x^{\prime}}+v_{2}\sigma_{2}\hat
{p}_{y^{\prime}}+v_{t}\sigma_{0}\hat{p}_{y^{\prime}},
\end{equation}
where $\hat{p}_{x^{\prime},y^{\prime}}$ are the momentum operators,
$\sigma_{1,2}$ and $\sigma_{0}$ are Pauli matrices and $2\times2$ identity
matrix, respectively. The anisotropic velocities are $v_{1}=0.86v_{F}$,
$v_{2}=0.69v_{F}$, and $v_{t}=0.32v_{F}$ with $v_{F}=10^{6}$ m/s. Throughout
this paper, we set $\hbar=v_{F}\equiv1$ to favor our dimensionless derivation
and calculations, and they can be used to define the length unit $l_{0}$\ and
the energy unit $\varepsilon_{0}$ through $\hbar v_{F}=l_{0}\varepsilon_{0}$,
e.g., $\varepsilon_{0}=0.66$ eV when $l_{0}=1$ nm. The eigenenergies and
eigenstates of $\hat{H}_{0}$ are, respectively,%

\begin{equation}
E_{\lambda,\mathbf{k}^{\prime}}=v_{t}k_{y^{\prime}}+\lambda\sqrt{v_{1}%
^{2}k_{x^{\prime}}^{2}+v_{2}^{2}k_{y^{\prime}}^{2}}, \label{ED}%
\end{equation}
and%

\begin{equation}
|\Psi_{\lambda,\mathbf{k}^{\prime}}(\mathbf{r}^{\prime})\rangle=\frac{1}%
{\sqrt{2}}%
\begin{bmatrix}
1\\
\exp(i\theta_{\mathbf{S}_{\lambda}(\mathbf{k}^{\prime})})
\end{bmatrix}
e^{i\mathbf{k}^{\prime}\cdot\mathbf{r}^{\prime}}. \label{WF}%
\end{equation}
Here, $\lambda=+$ ($\lambda=-$)\ denotes the conduction (valence) band,
$\mathbf{k}^{\prime}=(k_{x^{\prime}},k_{y^{\prime}})$ is the wave vector and
$\mathbf{r}^{\prime}=(x^{\prime},y^{\prime})$ is the position vector. In
particular, $\theta_{\mathbf{S}_{\lambda}(\mathbf{k}^{\prime})}$ is the
azimuthal angle of the in-plane pseudospin vector $\mathbf{S}_{\lambda
}(\mathbf{k}^{\prime})=\langle\Psi_{\lambda,\mathbf{k}^{\prime}}|(\sigma
_{1},\sigma_{2})|\Psi_{\lambda,\mathbf{k}^{\prime}}\rangle$ relative to
$x^{\prime}$-axis, which has the form\cite{PhysRevB.97.235440}: \ %

\begin{equation}
\mathbf{S}_{\lambda}(\mathbf{k}^{\prime})=\frac{\lambda\left(  v_{1}%
k_{x^{\prime}},v_{2}k_{y^{\prime}}\right)  }{\sqrt{v_{1}^{2}k_{x^{\prime}}%
^{2}+v_{2}^{2}k_{y^{\prime}}^{2}}}.
\end{equation}
The anisotropy and tilt of MDFs lead to the noncollinear feature of wave vector
and group velocity for a general state, which has profound modifications to
physical properties of 8-\textit{Pmmn} borophene comparing to those of
graphene, e.g, oblique Klein tunneling\cite{PhysRevB.97.235440}. And this
nonlinear feature can be clearly shown by the definition of group velocity
$\mathbf{v}_{\lambda}^{\prime}(\mathbf{k}^{\prime})\equiv\partial
_{\mathbf{k}^{\prime}}E_{\lambda,\mathbf{k}^{\prime}}=(v_{\lambda,x}^{\prime
},v_{\lambda,y}^{\prime})$\ with the components%

\begin{subequations}
\label{VG}%
\begin{align}
v_{\lambda,x}^{\prime}  &  =\frac{\lambda v_{1}^{2}k_{x^{\prime}}}{\sqrt
{v_{1}^{2}k_{x^{\prime}}^{2}+v_{2}^{2}k_{y^{\prime}}^{2}}},\\
v_{\lambda,y}^{\prime}  &  =v_{t}+\frac{\lambda v_{2}^{2}k_{y^{\prime}}}%
{\sqrt{v_{1}^{2}k_{x^{\prime}}^{2}+v_{2}^{2}k_{y^{\prime}}^{2}}}.%
\end{align}

The two-dimensional nature of 8-\textit{Pmmn} borophene is suitable to\ the
fabrication of the planar PNJ.\ One typical 8-\textit{Pmmn} borophene PNJ is
shown schematically by Fig. \ref{structure}(a). In Fig. \ref{structure}(a), we
use the normal (tangential)\ direction of junction interface to define $x$
($y$) axis of the Cartesian coordinate system $x-y$ which has a rotation angle
$\phi$ relative to the coordinate system $x^{\prime}-y^{\prime}$ for the
intrinsic Hamiltonian $\hat{H}_{0}$, so $\phi$ can be regarded as the junction
direction of $\phi$-junction. In the coordinate systems $x-y$ and $x^{\prime}-y^{\prime}$, an
arbitrary vector $\mathbf{A}$\ can be expressed as $\mathbf{A=(}A_{x}%
,A_{y}\mathbf{)}$ and $\mathbf{A}^{\prime}\mathbf{=(}A_{x^{\prime}%
},A_{y^{\prime}}\mathbf{)}$ and they are related to each other $\mathbf{A}%
^{\text{T}}=\mathbf{U(A}^{\prime})^{\text{T}}$ through the unitary matrix
\end{subequations}
\begin{equation}
\mathbf{U=}\left[
\begin{array}
[c]{cc}%
\cos\phi & \sin\phi\\
-\sin\phi & \cos\phi
\end{array}
\right]  . \label{UM}%
\end{equation}
The Hamiltonian of 8-\textit{Pmmn} borophene PNJ in Fig. \ref{structure}(a)
is
\begin{equation}
\hat{H}=(\hat{H}_{0}+V_{n})\Theta(-x)+(\hat{H}_{0}+V_{n})\Theta(x),
\end{equation}
where $V_{n}=-V_{0}$ ($V_{p}=V_{0}$) is the gate-induced scalar potential in
the \textit{n} (\textit{p}) region as shown by Fig. \ref{structure}(b)\ and no
loss of generality we assume $V_{0}>0$, and $\Theta(x)$ is the step function:
$\Theta(x)=1$ for $x>0$ and $\Theta(x)=0$ for $x<0$. The Fermi level $E_{F}$
determines the doping configuration $\varepsilon_{\alpha}$ with $\alpha=\mathit{n,p}$
in two regions through $\varepsilon_{\alpha}\equiv E_{F}-V_{\alpha}$, where a
positive (negative) doping level corresponding to electron (hole) doping, so
$\varepsilon_{\mathit{n}}>0$ and $\varepsilon_{\mathit{p}}<0$ for the PNJ. \ \

\subsection{Green's function}

The propagation of anisotropic and tilted MDFs in the PNJ structure
can be described by the corresponding propagator or GF
\begin{equation}
\mathbf{G}(E_{F},\mathbf{r}_{2},\mathbf{r}_{1})\equiv\langle\mathbf{r}%
_{2}|(E_{F}+i0^{+}-\hat{H})^{-1}|\mathbf{r}_{1}\rangle
\end{equation}
of the Hamiltonian $\hat{H}$. Usually, the position and energy dependence of
GF are not shown explicitly for brevity in our study, i.e., $\mathbf{G}%
\equiv\mathbf{G}(E_{F},\mathbf{r}_{2},\mathbf{r}_{1})$.\ For the quasi-one
dimensional systems such as the considered PNJ in this study, the GF can
be constructed through the on-shell spectral expansion\cite{ZhangPRB2017},
i.e., it is just the spectral expansion of the states on the Fermi surface
instead of all the eigenstates including on-shell and off-shell ones of the
system in the conventional spectral expansion method for the
GF\cite{SakuraiBook1994,GriffithsBook1995,CohenBook2005}.\ More importantly,
according to the on-shell spectral expansion\cite{ZhangPRB2017}, GF has a
physical transparent formalism which favors its analytical construction through
the intrinsic states and its subsequent scattering states by the scattering
mechanism in the quasi-one dimensional system, e.g., the junction interface of
the PNJ.

\begin{figure}[ptbh]
\includegraphics[width=1.0\columnwidth,clip]{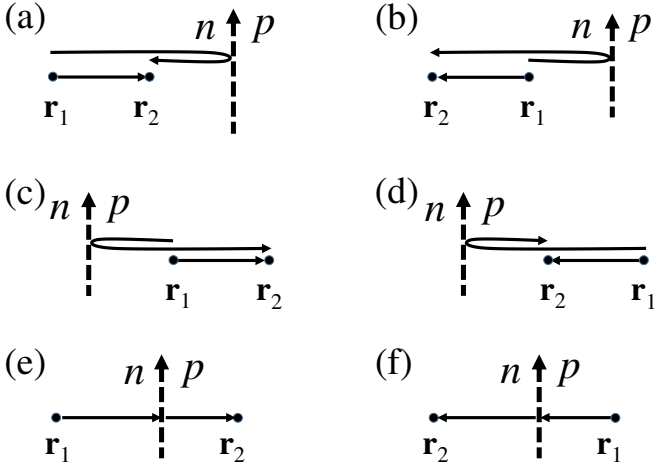}\caption{Piecewise
construction of Green's function (GF) $\mathbf{G}$ through intrinsic and
scattering states. (a)/(b) $\mathbf{G}$\ is the sum of intrinsic GF
$\mathbf{G}_{0}$ constructed through right-going/left-going intrinsic states
of the $\mathit{n}$ region), and extra GF constructed through the right-going
intrinsic states and its left-going reflection states by the PNJ interface.
(c)/(d) $\mathbf{G}$ is the sum of intrinsic GF $\mathbf{G}_{0}$ constructed
through right-going/left-going intrinsic states of the $\mathit{p}$ region and
extra GF constructed through the left-going intrinsic states and its
right-going reflection states by the PNJ interface. (e)/(f) $\mathbf{G}$ is
constructed through the right-going/left-going intrinsic states of the
$\mathit{n}$/$\mathit{p}$ region and its right-going/left-going transmission
states of the $\mathit{p}$/$\mathit{n}$ region. }%
\label{SGF}
\end{figure}

Fig. \ref{SGF} shows the piecewise construction of the GF $\mathbf{G}$ through
the intrinsic and scattering states in the PNJ structure. The intrinsic state
are the eigenstates of $\mathit{n}$ and $\mathit{p}$ regions in the PNJ while
the scattering states\ are the renormalized eigenstates by the scattering
coefficients. In Fig. \ref{SGF}, it is the interface of the PNJ leading to the
scattering of incident intrinsic states. Considering the different incident
states (maybe right-going or left-going)\ and the relative position of
$\mathbf{r}_{1}$ and $\mathbf{r}_{2}$, there are total 6 pieces for the GF as
shown by Fig. \ref{SGF}(a)-(f). To illustrate the operation of the
construction technique, no loss of generality, we assume the right-going
electron states incident from the left $\mathit{n}$ region of the PNJ, then
the resulting states can be expressed as:%

\begin{equation}
|\Psi(\mathbf{r})\rangle=\left\{
\begin{array}
[c]{ll}%
|\Psi_{\mathbf{k}_{\mathit{n},+}}(\mathbf{r})\rangle+r(k_{y})|\Psi
_{\mathbf{k}_{\mathit{n},-}}(\mathbf{r})\rangle & (x<0),\\
t(k_{y})|\Psi_{\mathbf{k}_{\mathit{p},+}}(\mathbf{r})\rangle & (x>0).
\end{array}
\right.  \label{RSS}%
\end{equation}
Here, $\mathbf{k}_{\alpha,\pm}=(k_{\alpha,\pm,x},k_{y})$ are the right-going
($+$)\ and left-going ($-$) wave vectors in the $\alpha$ region, and the
conservation of the tangential momentum $k_{y}$ has been used. For brevity, we omit the
subscript\ $\lambda$ of the eigenstates since its value can be fully
determined by the subscript $\alpha$ of the momentum through $\lambda
=\mathrm{sgn}(\epsilon_{\alpha})$, and we also use this simplification to the
pseudospin vector in the following. Since $k_{y}$ is conserved, the
next step is to derive $k_{\alpha,\pm,x}$. By using the unitary matrix (cf.
Eq. \ref{UM})\ for the vector transfromation\ between two coordinate systems,
we have%

\begin{subequations}
\label{CT}%
\begin{align}
k_{\alpha,\pm,x^{\prime}}  &  =k_{\alpha,\pm,x}\cos\phi-k_{y}\sin\phi,\\
k_{\alpha,\pm,y^{\prime}}  &  =k_{\alpha,\pm,x}\sin\phi+k_{y}\cos\phi.
\end{align}
Because $\mathbf{k}_{\alpha,\pm}^{\prime}=(k_{\alpha,\pm,x^{\prime}}%
,k_{\alpha,\pm,y^{\prime}})$ satisfies Eq. \ref{ED} of eigenenergies, we obtain%

\end{subequations}
\begin{equation}
Ak_{\alpha,\pm,x}^{2}+B_{\alpha}k_{\alpha,\pm,x}+C_{\alpha}=0, \label{QEOU}%
\end{equation}
where%

\begin{subequations}
\begin{align}
A  &  =\cos^{2}\phi+\gamma^{2}\sin^{2}\phi,\\
B_{\alpha}  &  =2\gamma_{2}\epsilon_{\alpha}\sin\phi-2(1-\gamma^{2})k_{y}%
\sin\phi\cos\phi,\\
C_{\alpha}  &  =k_{y}^{2}(\sin^{2}\phi+\gamma^{2}\cos^{2}\phi)+2\gamma
_{2}\epsilon_{\alpha}k_{y}\cos\phi-\epsilon_{\alpha}^{2}.
\end{align}
\end{subequations}
Here, $\gamma^{2}=\gamma_{1}^{2}-\gamma_{2}^{2}$ with $\gamma_{1}=v_{2}/v_{1}$
and $\gamma_{2}=v_{t}/v_{1}$, and $\epsilon_{\alpha}=\varepsilon_{\alpha
}/v_{1}$ with $\varepsilon_{\mathit{n}}=E_{F}+V_{0}$ and $\varepsilon
_{\mathit{p}}=E_{F}-V_{0}$. Eq. \ref{QEOU} is a quadratic equation with one
unknown which gives two roots for $k_{\alpha,\pm,x}$:

\begin{equation}
k_{\alpha,\pm,x}=\frac{-B_{\alpha}\pm\sqrt{B_{\alpha}^{2}-4AC_{\alpha}}}{2A}.
\label{AKX}%
\end{equation}
So $k_{\alpha,\pm,x}$ is derived, and then $\mathbf{k}_{\alpha,\pm}^{\prime}$
is given by Eq. \ref{CT}. And $r(k_{y})$ and $t(k_{y})$ are, respectively,
reflection and transmission coefficients, which had been derived for the PNJ
with a general junction direction\cite{PhysRevB.97.235440}:
\begin{subequations}
\label{SE}%
\begin{align}
r(k_{y})  &  =-\frac{\exp(i\theta_{\mathbf{S}({\mathbf{k}}_{\mathit{n},+}%
)})-\exp(i\theta_{{\mathbf{S}}({\mathbf{k}}_{\mathit{p},+})})}{\exp
(i\theta_{{\mathbf{S}}({\mathbf{k}}_{\mathit{n},-})})-\exp(i\theta
_{{\mathbf{S}}({\mathbf{k}}_{\mathit{p},+})})},\\
t(k_{y})  &  =\frac{\exp(i\theta_{{\mathbf{S}}({\mathbf{k}}_{\mathit{n},-}%
)})-\exp(i\theta_{\mathbf{S}({\mathbf{k}}_{\mathit{n},+})})}{\exp
(i\theta_{{\mathbf{S}}({\mathbf{k}}_{\mathit{n},-})})-\exp(i\theta
_{{\mathbf{S}}({\mathbf{k}}_{\mathit{p},+})})}.
\end{align}

Finally, GF can be constructed through the intrinsic and scattering state as
the components of Eq. \ref{RSS} for the resulting
states\cite{ZhangPRB2017,zhang2017}:
\end{subequations}
\begin{equation}
\mathbf{G}=\left\{
\begin{array}
[c]{ll}%
\mathbf{G}_{0}+\int dk_{y}\frac{r(k_{y})}{2\pi iv_{\mathit{n},x}}%
|\Psi_{\mathbf{k}_{\mathit{n},-}}(\mathbf{r}_{2})\rangle\langle\Psi
_{\mathbf{k}_{\mathit{n},+}}(\mathbf{r}_{1})| & (x_{2}<0),\\
\int dk_{y}\frac{t(k_{y})}{2\pi iv_{\mathit{n},x}}|\Psi_{\mathbf{k}%
_{\mathit{p},+}}(\mathbf{r}_{2})\rangle\langle\Psi_{\mathbf{k}_{\mathit{n},+}%
}(\mathbf{r}_{1})| & (x_{2}>0).
\end{array}
\right.  \label{FGF}%
\end{equation}
Here, $v_{\mathit{n},x}\equiv v_{\mathit{n},x}(\mathbf{k}_{n,+})$ is the $x$
component of the group velocity $\mathbf{v}_{\mathit{n}}(\mathbf{k}%
_{\mathit{n},+})$\ in the Cartesian coordinate system $x-y$. When $x_{2}<0$,
Eq. \ref{FGF} is the mathematical description of Fig. \ref{SGF}(a)/(b), which
is the sum of intrinsic GF $\mathbf{G}_{0}$ constructed through
right-going/left-going intrinsic states of the $\mathit{n}$ region (i.e., the
free GF $\mathbf{G}_{0}$)\ and extra GF constructed through the right-going
intrinsic states and its left-going reflection states by the PNJ interface.
The free GF $\mathbf{G}_{0}\equiv\mathbf{G}_{0}(E_{F},\mathbf{r}%
_{2},\mathbf{r}_{1})$ is constructed completely by the intrinsic states and
has the form:
\begin{equation}
\mathbf{G}_{0}=\int\frac{dk_{y}}{2\pi iv_{\mathit{n},x}}\left\{
\begin{array}
[c]{ll}%
|\Psi_{\mathbf{k}_{\mathit{n},+}}(\mathbf{r}_{2})\rangle\langle\Psi
_{\mathbf{k}_{\mathit{n},+}}(\mathbf{r}_{1})| & (x_{2}>x_{1}),\\
|\Psi_{\mathbf{k}_{\mathit{n},-}}(\mathbf{r}_{2})\rangle\langle\Psi
_{\mathbf{k}_{\mathit{n},-}}(\mathbf{r}_{1})| & (x_{2}<x_{1}).
\end{array}
\right.
\end{equation}
which is consistent with the result derived through the straightforward Fourier
transformation of the GF in momentum space\cite{zhangJMMM2019}. When $x_{2}>0$, Eq.
\ref{FGF} is the mathematical description of Fig. \ref{SGF}(e), which is
constructed through the right-going intrinsic states of the $\mathit{n}$
region and its right-going transmission states of the $\mathit{p}$ region.
Most importantly, when $x_{2}>0$, we have $|\Psi_{\mathbf{k}_{\mathit{p},+}%
}(\mathbf{r}_{2})\rangle\langle\Psi_{\mathbf{k}_{\mathit{n},+}}(\mathbf{r}%
_{1})|\propto\exp(i\varphi_{\mathit{np}})$ where
\begin{equation}
\varphi_{\mathit{np}}(k_{y})=\mathbf{k}_{\mathit{p},+}\cdot\mathbf{r}%
_{2}-\mathbf{k}_{\mathit{n,+}}\cdot\mathbf{r}_{1} \label{APH}%
\end{equation}
is the propagation phase accumulated through the negative refraction
of incident right-going intrinsic states in the $\mathit{n}$ region to become right-going transmission states in the $\mathit{p}$
region\cite{zhang2017}. Similarly, to assume the left-going electron states
incident from the right $\mathit{p}$ region of the PNJ, one can obtain the
mathematical description of Fig. \ref{SGF}(c), (d) and (f). Noting here, the
developed construction technique of GF is universal to the PNJ with a general
junction direction.

\section{ Results and discussions}

In this section, we present the numerical results to display the negative
refraction induced interference pattern of anisotropic and tilted MDFs across
the 8-\textit{Pmmn} borophene PNJ and discuss the underlying physics. No loss
of generality, we focus on the GF matrix element $\left\vert {G}_{11}%
(E_{F},\mathbf{r}_{2},\mathbf{r}_{1})\right\vert $, since the interference
pattern originates from the phase accumulation across the PNJ as shown in the
following and has no qualitative difference among different GF matrix elements.

\begin{figure}[ptbh]
\includegraphics[width=1.0\columnwidth,clip]{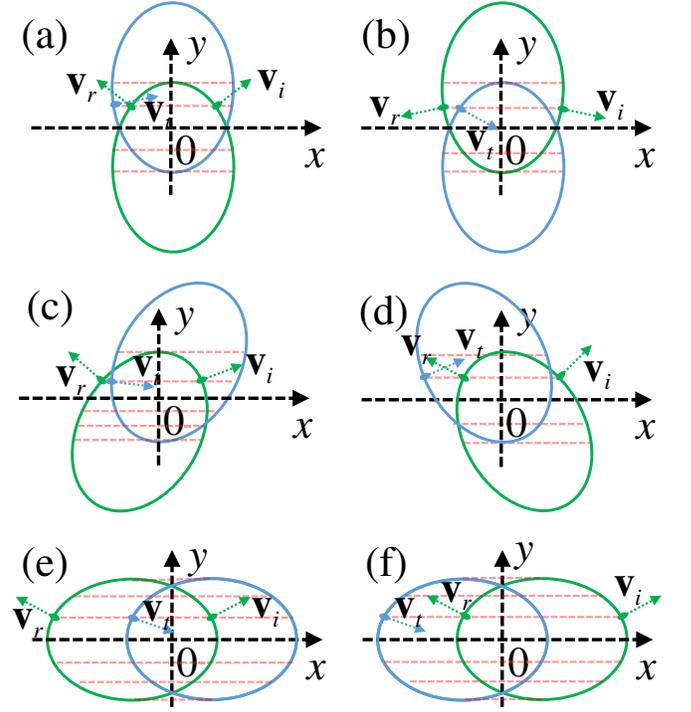}\caption{Dependence
of momentum matching on the junction direction $\phi$. (a) $\phi=0$, (b)
$\phi=\pi$, (c) $\phi=\pi/6$, (d) $\phi=-\pi/6$, (e) $\phi=\pi/2$, and (f)
$\phi=-\pi/2$. The available states on the electron (green) and hole (blue)
Fermi surfaces for the momentum matching subject to the conserved momentum
$k_{y}$ as shown by the red dashed lines. The colored lines with arrows represent the group velocities ${\bf v}_{i,r,t}$  for the incident, reflection and transmission states, which are nonlinear with the conserved momenta. Here, we use $\varepsilon
_{\mathit{n}}=\varepsilon_{\mathit{p}}=0.1$ as the doping configuration for
\textit{p-n} junctions.}%
\label{matching}%
\end{figure}

\begin{figure*}[ptbh]
\includegraphics[width=2.0\columnwidth]{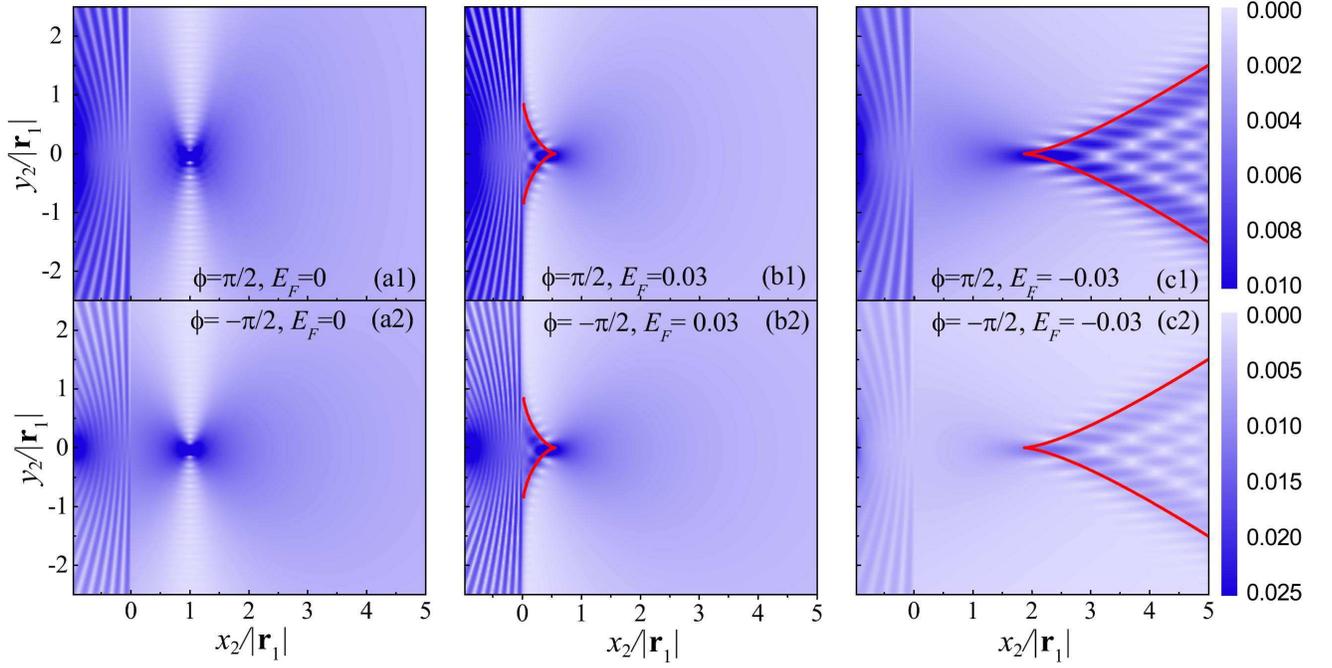}\caption{Veselago
focusing and normal caustics in\ \textit{p-n} junctions with junction
directions parallel to $x^{\prime}$ axis shown by $\left\vert {G}_{11}%
(E_{F},\mathbf{r}_{2},\mathbf{r}_{1})\right\vert $. Two different junction
directions (i.e., $\phi=\pm\pi/2$)\ and three different doping configurations
(i.e., $E_{F}=0$, $\pm0.03$) are considered. The red lines are the caustics
from analytical formula. Here, $\mathbf{r}_{1}=(-200,0)$ and $V_{0}=0.1$.}%
\label{Xcaustics}%
\end{figure*}

Previous to the detailed calculations, we firstly present the intuitive
physics for the interference pattern in 8-\textit{Pmmn} borophene PNJ. In
contrast to the isotropic MDFs across the graphene
PNJ\cite{CheianovScience2007}, the anisotropy and tilt of MDFs should lead to
the dependence of interference pattern on the junction direction of the
8-\textit{Pmmn} borophene PNJ. The junction direction dependence implies the distinct
momentum matching of states across borophene PNJ, then determines the
appearance and the effectiveness of interference pattern. Fig. \ref{matching}
shows the dependence of momentum matching on the junction direction $\phi$ of
the borophene PNJ. Six typical junction directions are plotted, i.e., $\phi
=0$, $\pi$, $\pm\pi/6$, and $\pm\pi/2$ as shown in Fig. \ref{matching}(a)-(f).
The available states on the electron (green) and hole (blue) Fermi surfaces
for the momentum matching subject to the conserved momentum $k_{y}$ (cf. red
dashed lines in Fig. \ref{matching}). In Fig. \ref{matching}, by assuming electron states incident
from the left \textit{n} region of the PNJ, we also plot the group velocities ${\bf v}_{i,r,t}$  for the right-going incident, left-going reflection and right-going transmission states. The noncollinear features between group velocities and conserved momenta are obvious. Comparing to the partial matching in
Fig. \ref{matching}(a)-(d), all states on the electron and hole Fermi surfaces
may contribute to interference pattern in Fig. \ref{matching}(e) and (f), then
lead to the effective interference. More importantly, in Fig. \ref{matching}%
(e) and (f), the electron and hole Fermi surfaces have the mirror symmetry
about the junction interface, so Veselago focusing is expected in these two
special junctions\cite{ZhangPRB2016}. As a representative interference
pattern, Veselago focusing are studied intensively and are understood
deeply\cite{CheianovScience2007,PhysRevLett.100.236801,MoghaddamPRL2010,SilveirinhaPRL2013,ZhaoPRL2013,MilovanovicJAP2015,ncomms15783,zhang2017,PhysRevB.95.214103}%
. Therefore, the interference pattern in the special junctions with $\phi
=\pm\pi/2$ being perpendicular to the tilt direction (namely, $y^{\prime}$
axis), can be as the starting point of our discussions. Then, we turn to the
junctions with $\phi=0$ and $\phi=\pi$ (being parallel to the tilt direction)
and the general junctions.

\subsection{$\phi=\pm\pi/2$: Veselago focusing and normal caustics}

We have picked out two special junctions perpendicular to the tilt direction (namely, $y^{\prime}$
axis) of MDFs in 8-\textit{Pmmn} borophene. In these two special junctions, we
plot the matrix element of GF $\left\vert {G}_{11}(E_{F},\mathbf{r}%
_{2},\mathbf{r}_{1})\right\vert $ in Fig. \ref{Xcaustics}. For two different
junction directions (i.e., $\phi=\pm\pi/2$), three different doping
configurations are considered (i.e., $E_{F}=0$, $\pm0.03$). For both special junctions, Veselago focusing as expected\cite{ZhangPRB2016} (normal
caustics very similar to that in graphene PNJ\cite{CheianovScience2007})
appears when $E_{F}=0$ ($E_{F}=\pm0.03$). For these two special junctions, we
can analytically reveal the underlying physics of Veselago focusing and
caustics. The tilt of MDFs in 8-\textit{Pmmn} borophene leads to the breaking
of mirror symmetry about the $x^{\prime}\ $axis, so two cases of junction
direction along $jx^{\prime}$ axis with $j=\pm$ are inequivalent (cf. momentum
matching in Fig. \ref{matching}(e) and (f)). For the junction direction along
$jx^{\prime}$ direction, i.e., $\phi=-j\pi/2$, $\mathbf{k}_{\alpha,\pm}%
^{j}=(k_{\alpha,\pm,x}^{j},k_{y})$ has the form (cf. Eq. \ref{AKX} for
momentum components $k_{\alpha,\pm,x}^{j}$):%

\begin{equation}
k_{\alpha,\pm,x}^{j}=\frac{1}{\gamma^{2}}(j\epsilon_{\alpha}\gamma_{2}%
\pm\lambda\sqrt{\epsilon_{\alpha}^{2}\gamma_{1}^{2}-\gamma^{2}k_{y}^{2}}).
\label{JK}%
\end{equation}
Here, $\lambda=\mathrm{sgn}(\epsilon_{\alpha})$ conforms to Eq. \ref{ED} of eigenenergies. To show the formation of interference pattern of
anisotropic and tilted MDFs across the PNJ, we examine the classical
trajectory determined by the phase accumulation $\varphi_{\mathit{np}}$ (cf.
Eq. \ref{APH}). The classical trajectory with specific $k_{y,c}$\ is
determined by $\partial_{k_{y}}\varphi_{\mathit{np}}^{j}(k_{y})|_{k_{y,c}}%
=0$\ as\cite{zhang2017}%

\begin{equation}
\partial_{k_{y}}k_{\mathit{p},+,x}^{j}x_{2}-\partial_{k_{y}}k_{\mathit{n}%
,+,x}^{j}x_{1}+(y_{2}-y_{1})=0.\label{CTE}%
\end{equation}
To differentiate Eq. \ref{JK}, one can obtain%

\begin{subequations}
\begin{align}
\partial_{k_{y}}k_{\mathit{n},+,x}^{j}  &  =-\frac{k_{y}}{\sqrt{\gamma_{1}%
^{2}\epsilon_{n}^{2}-\gamma^{2}k_{y}^{2}}},\\
\partial_{k_{y}}k_{\mathit{p},+,x}^{j}  &  =\frac{k_{y}}{\sqrt{\gamma_{1}%
^{2}\epsilon_{p}^{2}-\gamma^{2}k_{y}^{2}}}.
\end{align}
As a result, the classical trajectory with $k_{y,c}$ satisfies the equation:%

\end{subequations}
\begin{equation}
y_{2}-y_{1}=-(\frac{k_{y,c}x_{2}}{\sqrt{\gamma_{1}^{2}\epsilon_{p}^{2}%
-\gamma^{2}k_{y,c}^{2}}}+\frac{k_{y,c}x_{1}}{\sqrt{\gamma_{1}^{2}\epsilon
_{n}^{2}-\gamma^{2}k_{y,c}^{2}}}). \label{KYC}%
\end{equation}
Noting here, Eq. \ref{KYC} has no dependence on $j$, so it has the same form
for two special junctions with $\phi=\pm\pi/2$. And Eq. \ref{KYC} reproduces
the result for isotropic MDFs when $v_{t}=0$ and $v_{1}=v_{2}$, i.e.,
$\gamma=\gamma_{1}=1$\cite{PhysRevB.97.235440}.

\begin{figure*}[tbh]
\includegraphics[width=2.0\columnwidth]{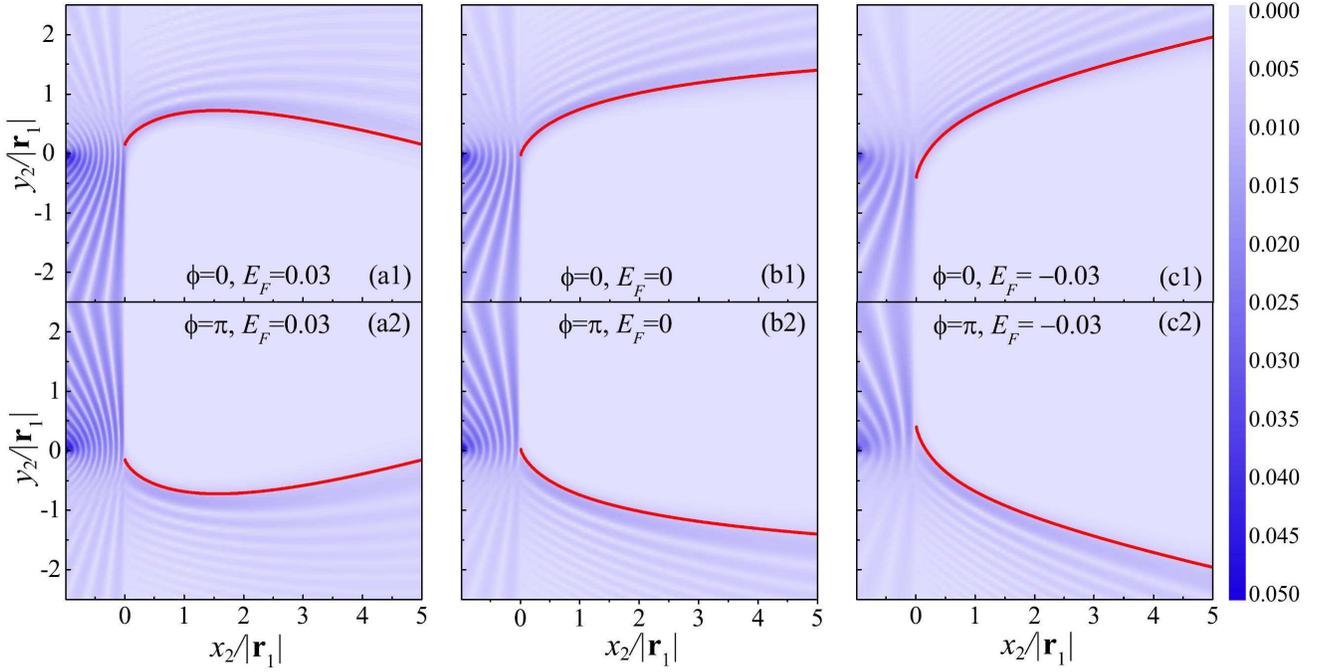}\caption{Anomalous
caustics in PNJs with junction directions parallel to $y^{\prime}$ axis (i.e.,
$\phi=0$, $\pi$) shown by $\left\vert {G}_{11}(E_{F},\mathbf{r}_{2}%
,\mathbf{r}_{1})\right\vert $. Three different Fermi levels ($E_{F}=0$,
$\pm0.03$) are considered. The red lines are the caustics from analytical
formula. Here, $\mathbf{r}_{1}=(-200,0)$ and $V_{0}=0.1$.}%
\label{Ycaustics1}%
\end{figure*}

\begin{figure}[htb]
\includegraphics[width=0.9\columnwidth]{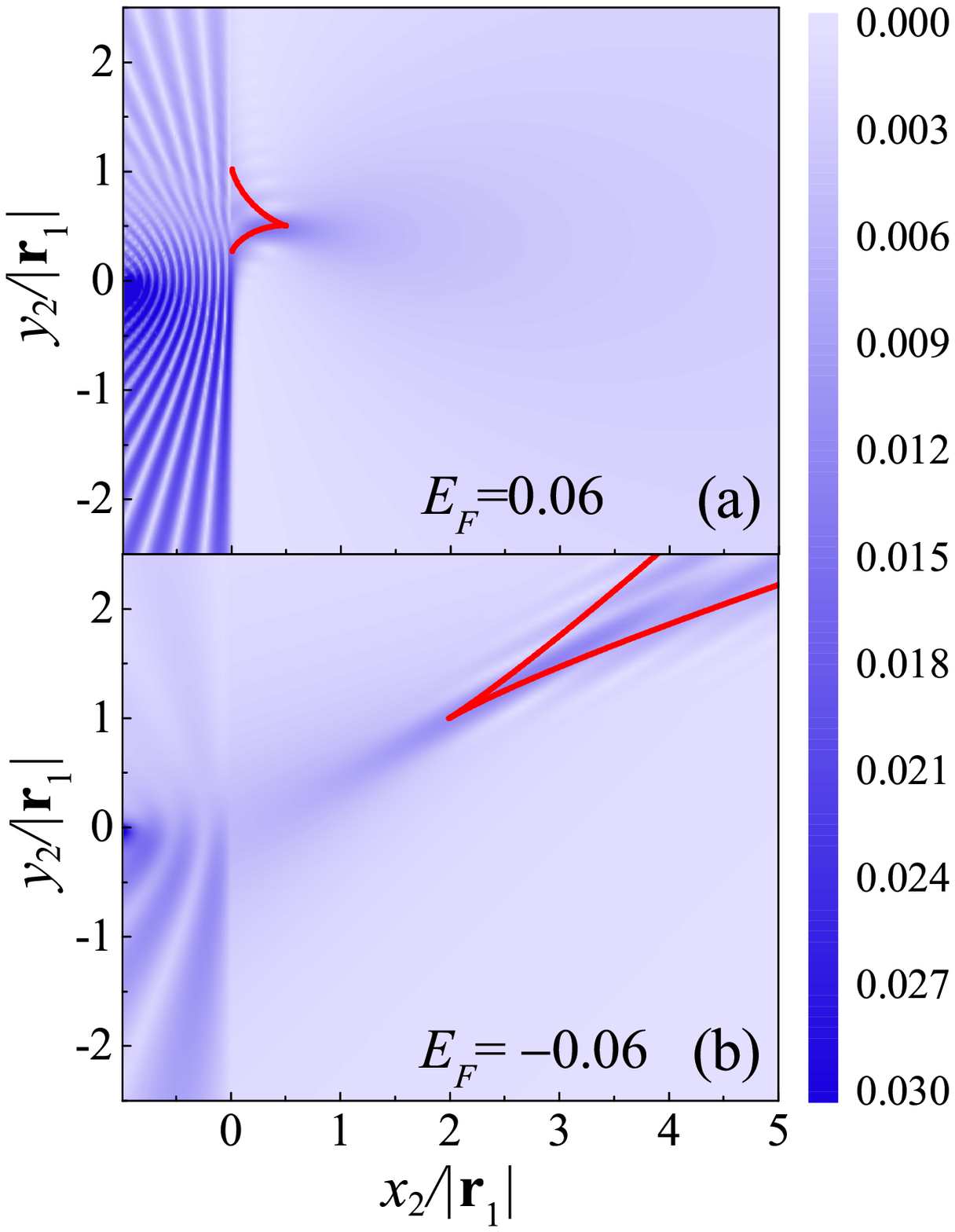}\caption{Same as Fig. \ref{Ycaustics1} except that two higher magnitude of Fermi levels ($E_{F}=\pm0.06$) are considered for $\phi=0$-junction.}%
\label{Ycaustics2}%
\end{figure}

\subsubsection{Symmetric doping: Veselago focusing}

To consider the symmetric doping, i.e., $E_{F}=0$ or $\epsilon_{n}%
=-\epsilon_{p}=V_{0}/v_{1}$, the classical trajectory is
\begin{equation}
y_{2}-y_{1}=-\frac{k_{y,c}(x_{2}+x_{1})}{\sqrt{\gamma_{1}^{2}\epsilon_{n}%
^{2}-\gamma^{2}k_{y,c}^{2}}}. \label{SVL}%
\end{equation}
Obviously, when $x_{1}+x_{2}=0$ and $y_{2}=y_{1}$, Eq. \ref{SVL}\ has no
dependence on the momentum of classical trajectory, this implies\ that all
classical trajectories from $\mathbf{r}_{1}=(x_{1},y_{1})$ converge to its
mirror point $\mathbf{r}_{2}=(-x_{1},y_{1})$ about the junction interface,
i.e., the appearance of Veselago focusing. Meanwhile, $k_{\mathit{n},+,x}%
^{j}=-k_{\mathit{p},+,x}^{j}$ and $\partial_{k_{y}}^{m}k_{\mathit{n},+,x}%
^{j}=-\partial_{k_{y}}^{m}k_{\mathit{p},+,x}^{j}$ with $m$ being a positive
integer, so the zero-order term $\varphi_{\mathit{np}}^{j}(k_{y,c}%
)=-k_{n,+,x}^{j}(x_{2}+x_{1})+k_{y,c}(y_{2}-y_{1})=0$ and the arbitrary order
derivative\ of propagation phase $\varphi^{j}_{\mathit{np}}$ on the classical
trajectory also vanishes identical to the case for isotropic MDFs in
graphene\cite{PhysRevB.97.235440}. However, noting that the different color
scales for the Veselago focusing and caustics on the upper and bottom rows of
Fig. \ref{Xcaustics}, implying the different focusing magnitude. This reflects
the inequivalence of two special junctions as mentioned previously, which has different density of
states for the focusing states.

\subsubsection{Asymmetric doping: normal caustics}

To consider the asymmetric doping configuration (i.e., $E_{F}\neq0$), Veselago
focusing disappears, but the classical trajectories can form another kind of
interference pattern (i.e., caustics) if the quadratic term$\ $of
$\varphi_{\mathit{np}}(k_{y})$ vanishes\cite{PhysRevB.97.235440}. Using Eq.
\ref{JK}, the vanishing quadratic term $\partial_{k_{y}^{2}}^{2}%
\varphi_{\mathit{np}}(k_{y})|_{k_{y,c}}=0$\ leads to%

\begin{equation}
\frac{\gamma_{1}^{2}\epsilon_{p}^{2}}{\left(  \gamma_{1}^{2}\epsilon_{p}%
^{2}-\gamma^{2}k_{y,c}^{2}\right)  ^{\frac{3}{2}}}x_{2}+\frac{\gamma_{1}%
^{2}\epsilon_{n}^{2}}{\left(  \gamma_{1}^{2}\epsilon_{n}^{2}-\gamma^{2}%
k_{y,c}^{2}\right)  ^{\frac{3}{2}}}x_{1}=0.
\end{equation}
To solve the above equation, we derive%

\begin{equation}
k_{y,c}^{2}=\frac{\gamma_{1}^{2}(c\epsilon_{n}^{2}-\epsilon_{p}^{2}%
)}{(c-1)\gamma^{2}}%
\end{equation}
where $c=n^{2}(x_{2}/x_{\text{cump}})^{2/3}$ with $n=-\epsilon_{p}%
/\epsilon_{n}$ and $x_{\text{cump}}=-nx_{1}$. As a result, the classical
trajectory is%

\begin{subequations}
\label{CAUS}%
\begin{align}
y_{2}-y_{1}  &  =\pm\frac{(x_{2}^{2/3}-x_{\text{cusp}}^{2/3})^{3/2}}%
{\gamma\sqrt{(n^{2}-1)}}\text{, for }n>1\text{ \& }x_{2}>x_{\text{cusp}},\\
y_{2}-y_{1}  &  =\pm\frac{(x_{\text{cusp}}^{2/3}-x_{2}^{2/3})^{3/2}}%
{\gamma\sqrt{1-n^{2}}}\text{, for }n<1\text{ \& }x_{2}<x_{\text{cusp}}.
\end{align}
Eq. \ref{CAUS} reproduces the result for isotropic MDFs
\cite{PhysRevB.97.235440} when $\gamma=\gamma_{1}=1$, and it just indicates
the normal caustics. Due to the anisotropy ($v_{1}\neq v_{2}$) and tilt
($v_{t}\neq 0$) of MDFs in 8-\textit{Pmmn} borophene, $\gamma\neq1$ leads to
the modification of caustics. From a different perspective, this implies the
tunability of caustics by changing the anisotropy and tilt of MDFs. In
addition, the analytical Eq. \ref{CAUS} is used to plot the red lines in Fig.
\ref{Xcaustics}, and they are very consistent with the numerical results.

\subsection{$\phi=0$ and $\pi$: anomalous caustics}

Fig. \ref{Ycaustics1} shows $\left\vert {G}_{11}(E_{F},\mathbf{r}%
_{2},\mathbf{r}_{1})\right\vert$ in PNJs with junction directions (i.e., $\phi=0$, $\pi$) parallel to
tilt direction (namely $y^\prime$ axis). In Fig. \ref{Ycaustics1}, an outstanding feature is that the upper and bottom panels have the mirror symmetry about the $x$ axis. This comes from one hidden symmetry: To perform inversion operations $y\rightarrow -y$ and $k_y\rightarrow -k_{y}$ for $\phi$-junction, its momentum matching becomes the case of $(\pi-\phi)$-junction (cf. Fig. \ref{matching}). Using this hidden symmetry, one can also explain the mirror symmetry of interference pattern about the $x$ axis in Fig. \ref{Xcaustics} since $\pi/2$-junction or $-\pi/2$-junction is symmetrically related to itself.

Comparing to Fig. \ref{Xcaustics}
for special junctions, Veselago focusing does not exist and caustics becomes
anomalous in Fig. \ref{Ycaustics1}, i.e., anomalous caustics. The MDFs in 8-\textit{Pmmn} borophene have two features: anisotropy ($v_1\neq v_2$) and tilt ($v_t\neq0$). To assume $v_t=0$, the electron and hole Fermi surfaces have the mirror symmetry about the PNJ interface when $E_F=0$, so Veselago focusing exists\cite{ZhangPRB2016,PhysRevB.97.205437}. The absence of Veselago focusing in Fig. \ref{Ycaustics1} is due to the breaking of mirror symmetry by the finite tilt (cf. Fig. \ref{matching}(a) and (b)). Nevertheless, the anisotropy and tilt both contribute to the caustics, even in the special junctions (cf. Eq. \ref{CAUS}). The $\phi$-dependent momentum matching further introduces the dependence of the anomalous caustics on the junction direction. For the
junction direction along $jy^{\prime}$ direction, we have $\mathbf{k}%
_{\alpha,\pm}^{j}=(k_{\alpha,\pm,x}^{j},k_{y})$ and $k_{\alpha,\pm,x}^{j}%
$\ are given by Eq. \ref{AKX}:%

\end{subequations}
\begin{equation}
k_{\alpha,\pm,x}^{j}=\pm\lambda\sqrt{\epsilon_{\alpha}^{2}-k_{y}^{2}\gamma
^{2}-2j\gamma_{2}\epsilon_{\alpha}k_{y}},
\end{equation}
which implies
\begin{subequations}
\begin{align}
\partial_{k_{y}}k_{\mathit{n},+,x}^{j} &  =-\frac{k_{y}\gamma^{2}+j\gamma
_{2}\epsilon_{n}}{\sqrt{-\gamma^{2}k_{y}^{2}+\epsilon_{n}^{2}-2j\gamma
_{2}\epsilon_{n}k_{y}}},\\
\partial_{k_{y}}k_{\mathit{p},+,x}^{j} &  =\frac{k_{y}\gamma^{2}+j\gamma
_{2}\epsilon_{p}}{\sqrt{-\gamma^{2}k_{y}^{2}+\epsilon_{p}^{2}-2j\gamma
_{2}\epsilon_{p}k_{y}}}.
\end{align}
\label{1DKX}
Using Eq. \ref{APH}, the vanishing quadratic term $\partial_{k_{y}^{2}}%
^{2}\varphi_{\mathit{np}}(k_{y})|_{k_{y,c}}=0$\ leads to%

\end{subequations}
\begin{equation}
\frac{\epsilon_{p}^{2}x_{2}}{\left(  -\gamma^{2}k_{y}^{2}+\epsilon_{p}%
^{2}-2j\gamma_{2}\epsilon_{p}k_{y}\right)  ^{\frac{3}{2}}}=\frac{-\epsilon
_{n}^{2}x_{1}}{\left(  -\gamma^{2}k_{y}^{2}+\epsilon_{n}^{2}-2j\gamma
_{2}\epsilon_{n}k_{y}\right)  ^{\frac{3}{2}}}.
\end{equation}
\bigskip The solution of the above equation is   %

\begin{equation}
k_{y,c}^{j}=\frac{-j\gamma_{2}(c\epsilon_{n}-\epsilon_{p})\pm\sqrt{(\gamma
_{2}^{2}c\epsilon_{n}-\epsilon_{p})^{2}-\gamma^{2}(c-1)(\epsilon_{p}%
^{2}-c\epsilon_{n}^{2})}}{(c-1)\gamma^{2}}.\label{XKYC}%
\end{equation}
Substituting Eqs. \ref{1DKX} and \ref{XKYC}\ into Eq. \ref{CTE}, the anomalous
caustics can be determined and is shown by the red lines in Fig.
\ref{Ycaustics1}. The analytical and numerical results agree
very well with each other.

In addition, Fig. \ref{Ycaustics1} also shows the dependence of anomalous caustics on the doping configuration. With decreasing $E_F$ from left to right on the upper (bottom) row, the anomalous caustics goes upward (downward). This implies the tunable anomalous caustics by doping configuration. It attracts us to further increase the magnitude of $E_F$ to examine the interference pattern. Fig. \ref{Ycaustics2} shows the interference pattern of $\left\vert {G}_{11}(E_{F},\mathbf{r}%
_{2},\mathbf{r}_{1})\right\vert $ same as Fig. \ref{Ycaustics1} except that two higher magnitude of Fermi levels ($E_{F}=\pm0.06$) are considered for ($\phi=0$)-junction. On the caustics, Fig. \ref{Ycaustics1} has a single branch, whereas there are two branches in Fig. \ref{Ycaustics2}. In the sense of the same branch number for caustics in Fig. \ref{Xcaustics} and Fig. \ref{Ycaustics2}, the anomalous feature is suppressed by the high value of $E_F$. And there should be one critical $E_F$ at which the anomalous caustics changes its branch number.  In all, the anomalous caustics has a strong dependence on doping configuration and exhibits rich patterns.

\subsection{General $\phi$: anomalous caustics}

\begin{figure*}[htb]
\includegraphics[width=2.0\columnwidth]{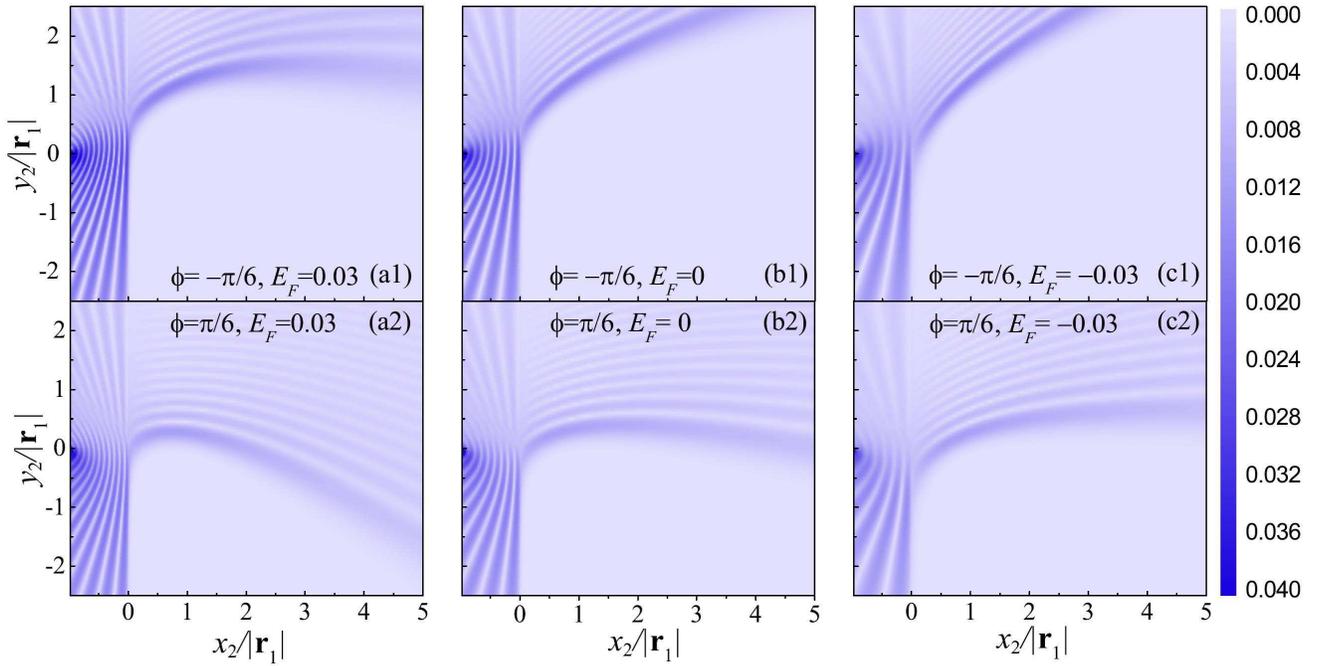}\caption{Anomalous
caustics in the PNJ with general junction directions shown by $\left\vert {G}_{11}(E_{F},\mathbf{r}_{2}%
,\mathbf{r}_{1})\right\vert$. Two different junction
directions (i.e., $\phi=\pm\pi/6$)\ and three different Fermi levels (i.e.,
$E_{F}=0$, $\pm0.03$) are considered.}
\label{Gcaustics}%
\end{figure*}

In general cases, the junction directions $\phi$ of PNJs are neither perpendicular nor parallel to the tilt direction of MDFs in 8-\textit{Pmmn}
borophene. It is rather difficult to examine the transport properties of these general $\phi$-junctions, because their low symmetry requires large unit cell to perform numerical simulation, e.g., through the recursive GF based on the lattice Hamiltonian\cite{PhysRevB.80.155406}. By using the developed construction technique of GF in Sec. IIB, the transport problem in general $\phi$-junctions can be solved efficiently. Fig. \ref{Gcaustics} shows $\left\vert {G}_{11}(E_{F},\mathbf{r}_{2}%
,\mathbf{r}_{1})\right\vert$ by considering two different junction
directions (i.e., $\phi=\pm\pi/6$) and three different Fermi levels (i.e.,
$E_{F}=0$, $\pm0.03$). In light of the mirror symmetry about the $x$ axis for $\phi$-junction and $(\pi-\phi)$-junction, one only needs to explicitly discuss the junction directions $\phi\in[-\pi/2,\pi/2]$, so $\phi=\pm \pi/6$ are typical enough. Fig. \ref{Gcaustics} is very similar to Fig. \ref{Ycaustics1}, featured by the anomalous caustics. And with decreasing $E_F$ from left to right on the upper or bottom row, the anomalous caustics goes upward. Therefore, the anomalous caustics depends strongly on the junction directions and doping configuration, favoring its tunability.


\section{Conclusions}

In summary, we investigate the negative refraction induced interference of anisotropic and tilted MDFs across 8-\textit{Pmmn} borophene
PNJs. Because of the anisotropy and tilt, the calculation of GF of the PNJ with an arbitrary junction direction is needed, for which we develop an  effective construction technique. The developed construction technique of GF can be generalized to study the transport of other novel MDFs, e.g., type-II Dirac fermions\cite{lu2016,PhysRevX.9.031010}, and semi-Dirac fermions \cite{PhysRevLett.103.016402}. Comparing to the seminal work\cite{CheianovScience2007}, here we focus negative refraction induced Veselago focusing and caustics. If the junction direction is perpendicular to the tilt
direction of MDFs, Veselago focusing or normal caustics appears resting on the doping configuration of the PNJ. More importantly, to the other junction, we discover one new phenomenon, i.e., anomalous caustics, which can be manipulated by junction direction and doping configuration. The interference pattern is unique to anisotropic and tilted MDFs, which is demonstrated numerically and analytically. This model study is generally applicable to the ideal \textit{p-n} junctions with perfect border, while the imperfect border (e.g., the disorder-induced losses and smooth junctions) will lead to the diffusive scattering and reduce the transmission of electrons across the electrons. To implement the Dirac fermions into electron optics, people make a great effort to improve the experimental fabrication of ideal \textit{p-n} junction. For example, very ideal atomically sharp \textit{p-n} junction have been created \cite{PhysRevB.97.045413,acsnano.8b09575,C9NR02029B}. In addition, with the rapid experimental advance of borophene \cite{MannixS1513,FengNC2016,PhysRevLett.118.096401,Wang2019} and the demonstration of negation refraction \cite{LeeNatPhys2015,ChenScience2016,PhysRevB.100.041401}, we expect the anomalous caustics to be observable in the near future, then this study make novel MDFs (e.g., in 8-\textit{Pmmn} borophene) promising to engineer Dirac electron optics devices.

\section*{Acknowledgements}

This work was supported by the National Key R$\&$D Program of China (Grant No.
2017YFA0303400), the NSFC (Grants No. 11504018, and No. 11774021), the MOST of
China (Grants No. 2014CB848700), and the NSFC program for \textquotedblleft
Scientific Research Center\textquotedblright\ (Grant No. U1530401). We
acknowledge the computational support from the Beijing Computational Science
Research Center (CSRC).


\begin{thebibliography}{55}
\expandafter\ifx\csname natexlab\endcsname\relax\def\natexlab#1{#1}\fi
\expandafter\ifx\csname bibnamefont\endcsname\relax
  \def\bibnamefont#1{#1}\fi
\expandafter\ifx\csname bibfnamefont\endcsname\relax
  \def\bibfnamefont#1{#1}\fi
\expandafter\ifx\csname citenamefont\endcsname\relax
  \def\citenamefont#1{#1}\fi
\expandafter\ifx\csname url\endcsname\relax
  \def\url#1{\texttt{#1}}\fi
\expandafter\ifx\csname urlprefix\endcsname\relax\def\urlprefix{URL }\fi
\providecommand{\bibinfo}[2]{#2}
\providecommand{\eprint}[2][]{\url{#2}}

\bibitem[{\citenamefont{Pendry}(2000)}]{PendryPRL2000}
\bibinfo{author}{\bibfnamefont{J.~B.} \bibnamefont{Pendry}},
  \bibinfo{journal}{Phys. Rev. Lett.} \textbf{\bibinfo{volume}{85}},
  \bibinfo{pages}{3966} (\bibinfo{year}{2000}).

\bibitem[{\citenamefont{Smith et~al.}(2004)\citenamefont{Smith, Pendry, and
  Wiltshire}}]{Smith788}
\bibinfo{author}{\bibfnamefont{D.~R.} \bibnamefont{Smith}},
  \bibinfo{author}{\bibfnamefont{J.~B.} \bibnamefont{Pendry}},
  \bibnamefont{and} \bibinfo{author}{\bibfnamefont{M.~C.~K.}
  \bibnamefont{Wiltshire}}, \bibinfo{journal}{Science}
  \textbf{\bibinfo{volume}{305}}, \bibinfo{pages}{788} (\bibinfo{year}{2004}).

\bibitem[{\citenamefont{Chen et~al.}(2016{\natexlab{a}})\citenamefont{Chen,
  Taylor, and Yu}}]{Chen2016}
\bibinfo{author}{\bibfnamefont{H.-T.} \bibnamefont{Chen}},
  \bibinfo{author}{\bibfnamefont{A.~J.} \bibnamefont{Taylor}},
  \bibnamefont{and} \bibinfo{author}{\bibfnamefont{N.}~\bibnamefont{Yu}},
  \bibinfo{journal}{Reports on Progress in Physics}
  \textbf{\bibinfo{volume}{79}}, \bibinfo{pages}{076401}
  (\bibinfo{year}{2016}{\natexlab{a}}).

\bibitem[{\citenamefont{Castro~Neto et~al.}(2009)\citenamefont{Castro~Neto,
  Guinea, Peres, Novoselov, and Geim}}]{CastroRMP2009}
\bibinfo{author}{\bibfnamefont{A.~H.} \bibnamefont{Castro~Neto}},
  \bibinfo{author}{\bibfnamefont{F.}~\bibnamefont{Guinea}},
  \bibinfo{author}{\bibfnamefont{N.~M.~R.} \bibnamefont{Peres}},
  \bibinfo{author}{\bibfnamefont{K.~S.} \bibnamefont{Novoselov}},
  \bibnamefont{and} \bibinfo{author}{\bibfnamefont{A.~K.} \bibnamefont{Geim}},
  \bibinfo{journal}{Rev. Mod. Phys.} \textbf{\bibinfo{volume}{81}},
  \bibinfo{pages}{109} (\bibinfo{year}{2009}).

\bibitem[{\citenamefont{Low}(2012)}]{Low2012}
\bibinfo{author}{\bibfnamefont{T.}~\bibnamefont{Low}},
  \emph{\bibinfo{title}{Graphene pn Junction: Electronic Transport and
  Devices}} (\bibinfo{publisher}{Springer Berlin Heidelberg},
  \bibinfo{address}{Berlin, Heidelberg}, \bibinfo{year}{2012}), pp.
  \bibinfo{pages}{467--508}.

\bibitem[{\citenamefont{Frisenda et~al.}(2018)\citenamefont{Frisenda,
  Molina-Mendoza, Mueller, Castellanos-Gomez, and van~der Zant}}]{C7CS00880E}
\bibinfo{author}{\bibfnamefont{R.}~\bibnamefont{Frisenda}},
  \bibinfo{author}{\bibfnamefont{A.~J.} \bibnamefont{Molina-Mendoza}},
  \bibinfo{author}{\bibfnamefont{T.}~\bibnamefont{Mueller}},
  \bibinfo{author}{\bibfnamefont{A.}~\bibnamefont{Castellanos-Gomez}},
  \bibnamefont{and} \bibinfo{author}{\bibfnamefont{H.~S.~J.}
  \bibnamefont{van~der Zant}}, \bibinfo{journal}{Chem. Soc. Rev.}
  \textbf{\bibinfo{volume}{47}}, \bibinfo{pages}{3339} (\bibinfo{year}{2018}).

\bibitem[{\citenamefont{Cheianov et~al.}(2007)\citenamefont{Cheianov, Fal'ko,
  and Altshuler}}]{CheianovScience2007}
\bibinfo{author}{\bibfnamefont{V.~V.} \bibnamefont{Cheianov}},
  \bibinfo{author}{\bibfnamefont{V.}~\bibnamefont{Fal'ko}}, \bibnamefont{and}
  \bibinfo{author}{\bibfnamefont{B.~L.} \bibnamefont{Altshuler}},
  \bibinfo{journal}{Science} \textbf{\bibinfo{volume}{315}},
  \bibinfo{pages}{1252} (\bibinfo{year}{2007}).

\bibitem[{\citenamefont{Lee et~al.}(2015)\citenamefont{Lee, Park, and
  Lee}}]{LeeNatPhys2015}
\bibinfo{author}{\bibfnamefont{G.-H.} \bibnamefont{Lee}},
  \bibinfo{author}{\bibfnamefont{G.-H.} \bibnamefont{Park}}, \bibnamefont{and}
  \bibinfo{author}{\bibfnamefont{H.-J.} \bibnamefont{Lee}},
  \bibinfo{journal}{Nat. Phys.} \textbf{\bibinfo{volume}{11}},
  \bibinfo{pages}{925} (\bibinfo{year}{2015}).

\bibitem[{\citenamefont{Chen et~al.}(2016{\natexlab{b}})\citenamefont{Chen,
  Han, Elahi, Habib, Wang, Wen, Gao, Taniguchi, Watanabe, Hone
  et~al.}}]{ChenScience2016}
\bibinfo{author}{\bibfnamefont{S.}~\bibnamefont{Chen}},
  \bibinfo{author}{\bibfnamefont{Z.}~\bibnamefont{Han}},
  \bibinfo{author}{\bibfnamefont{M.~M.} \bibnamefont{Elahi}},
  \bibinfo{author}{\bibfnamefont{K.~M.~M.} \bibnamefont{Habib}},
  \bibinfo{author}{\bibfnamefont{L.}~\bibnamefont{Wang}},
  \bibinfo{author}{\bibfnamefont{B.}~\bibnamefont{Wen}},
  \bibinfo{author}{\bibfnamefont{Y.}~\bibnamefont{Gao}},
  \bibinfo{author}{\bibfnamefont{T.}~\bibnamefont{Taniguchi}},
  \bibinfo{author}{\bibfnamefont{K.}~\bibnamefont{Watanabe}},
  \bibinfo{author}{\bibfnamefont{J.}~\bibnamefont{Hone}}, \bibnamefont{et~al.},
  \bibinfo{journal}{Science} \textbf{\bibinfo{volume}{353}},
  \bibinfo{pages}{1522} (\bibinfo{year}{2016}{\natexlab{b}}).

\bibitem[{\citenamefont{Garcia-Pomar et~al.}(2008)\citenamefont{Garcia-Pomar,
  Cortijo, and Nieto-Vesperinas}}]{PhysRevLett.100.236801}
\bibinfo{author}{\bibfnamefont{J.~L.} \bibnamefont{Garcia-Pomar}},
  \bibinfo{author}{\bibfnamefont{A.}~\bibnamefont{Cortijo}}, \bibnamefont{and}
  \bibinfo{author}{\bibfnamefont{M.}~\bibnamefont{Nieto-Vesperinas}},
  \bibinfo{journal}{Phys. Rev. Lett.} \textbf{\bibinfo{volume}{100}},
  \bibinfo{pages}{236801} (\bibinfo{year}{2008}).

\bibitem[{\citenamefont{Moghaddam and Zareyan}(2010)}]{MoghaddamPRL2010}
\bibinfo{author}{\bibfnamefont{A.~G.} \bibnamefont{Moghaddam}}
  \bibnamefont{and} \bibinfo{author}{\bibfnamefont{M.}~\bibnamefont{Zareyan}},
  \bibinfo{journal}{Phys. Rev. Lett.} \textbf{\bibinfo{volume}{105}},
  \bibinfo{pages}{146803} (\bibinfo{year}{2010}).

\bibitem[{\citenamefont{Silveirinha and Engheta}(2013)}]{SilveirinhaPRL2013}
\bibinfo{author}{\bibfnamefont{M.~G.} \bibnamefont{Silveirinha}}
  \bibnamefont{and} \bibinfo{author}{\bibfnamefont{N.}~\bibnamefont{Engheta}},
  \bibinfo{journal}{Phys. Rev. Lett.} \textbf{\bibinfo{volume}{110}},
  \bibinfo{pages}{213902} (\bibinfo{year}{2013}).

\bibitem[{\citenamefont{Zhao et~al.}(2013)\citenamefont{Zhao, Tang, Gu, and
  Duan}}]{ZhaoPRL2013}
\bibinfo{author}{\bibfnamefont{L.}~\bibnamefont{Zhao}},
  \bibinfo{author}{\bibfnamefont{P.}~\bibnamefont{Tang}},
  \bibinfo{author}{\bibfnamefont{B.-L.} \bibnamefont{Gu}}, \bibnamefont{and}
  \bibinfo{author}{\bibfnamefont{W.}~\bibnamefont{Duan}},
  \bibinfo{journal}{Phys. Rev. Lett.} \textbf{\bibinfo{volume}{111}},
  \bibinfo{pages}{116601} (\bibinfo{year}{2013}).

\bibitem[{\citenamefont{Milovanovic et~al.}(2015)\citenamefont{Milovanovic,
  Moldovan, and Peeters}}]{MilovanovicJAP2015}
\bibinfo{author}{\bibfnamefont{S.~P.} \bibnamefont{Milovanovic}},
  \bibinfo{author}{\bibfnamefont{D.}~\bibnamefont{Moldovan}}, \bibnamefont{and}
  \bibinfo{author}{\bibfnamefont{F.~M.} \bibnamefont{Peeters}},
  \bibinfo{journal}{J. Appl. Phys.} \textbf{\bibinfo{volume}{118}},
  \bibinfo{pages}{154308} (\bibinfo{year}{2015}).

\bibitem[{\citenamefont{B{\o}ggild et~al.}(2017)\citenamefont{B{\o}ggild,
  Caridad, Stampfer, Calogero, Papior, and Brandbyge}}]{ncomms15783}
\bibinfo{author}{\bibfnamefont{P.}~\bibnamefont{B{\o}ggild}},
  \bibinfo{author}{\bibfnamefont{J.~M.} \bibnamefont{Caridad}},
  \bibinfo{author}{\bibfnamefont{C.}~\bibnamefont{Stampfer}},
  \bibinfo{author}{\bibfnamefont{G.}~\bibnamefont{Calogero}},
  \bibinfo{author}{\bibfnamefont{N.~R.} \bibnamefont{Papior}},
  \bibnamefont{and}
  \bibinfo{author}{\bibfnamefont{M.}~\bibnamefont{Brandbyge}},
  \bibinfo{journal}{Nature Communications} \textbf{\bibinfo{volume}{8}},
  \bibinfo{pages}{15783} (\bibinfo{year}{2017}).

\bibitem[{\citenamefont{Zhang et~al.}(2017{\natexlab{a}})\citenamefont{Zhang,
  Zhu, Yang, and Chang}}]{zhang2017}
\bibinfo{author}{\bibfnamefont{S.-H.} \bibnamefont{Zhang}},
  \bibinfo{author}{\bibfnamefont{J.-J.} \bibnamefont{Zhu}},
  \bibinfo{author}{\bibfnamefont{W.}~\bibnamefont{Yang}}, \bibnamefont{and}
  \bibinfo{author}{\bibfnamefont{K.}~\bibnamefont{Chang}}, \bibinfo{journal}{2D
  Materials} \textbf{\bibinfo{volume}{4}}, \bibinfo{pages}{035005}
  (\bibinfo{year}{2017}{\natexlab{a}}).

\bibitem[{\citenamefont{Hills et~al.}(2017)\citenamefont{Hills, Kusmartseva,
  and Kusmartsev}}]{PhysRevB.95.214103}
\bibinfo{author}{\bibfnamefont{R.~D.~Y.} \bibnamefont{Hills}},
  \bibinfo{author}{\bibfnamefont{A.}~\bibnamefont{Kusmartseva}},
  \bibnamefont{and} \bibinfo{author}{\bibfnamefont{F.~V.}
  \bibnamefont{Kusmartsev}}, \bibinfo{journal}{Phys. Rev. B}
  \textbf{\bibinfo{volume}{95}}, \bibinfo{pages}{214103}
  (\bibinfo{year}{2017}).

\bibitem[{\citenamefont{Zhang et~al.}(2018)\citenamefont{Zhang, Yang, and
  Peeters}}]{PhysRevB.97.205437}
\bibinfo{author}{\bibfnamefont{S.-H.} \bibnamefont{Zhang}},
  \bibinfo{author}{\bibfnamefont{W.}~\bibnamefont{Yang}}, \bibnamefont{and}
  \bibinfo{author}{\bibfnamefont{F.~M.} \bibnamefont{Peeters}},
  \bibinfo{journal}{Phys. Rev. B} \textbf{\bibinfo{volume}{97}},
  \bibinfo{pages}{205437} (\bibinfo{year}{2018}).

\bibitem[{\citenamefont{Betancur-Ocampo}(2018)}]{PhysRevB.98.205421}
\bibinfo{author}{\bibfnamefont{Y.}~\bibnamefont{Betancur-Ocampo}},
  \bibinfo{journal}{Phys. Rev. B} \textbf{\bibinfo{volume}{98}},
  \bibinfo{pages}{205421} (\bibinfo{year}{2018}).

\bibitem[{\citenamefont{Prabhakar et~al.}(2019)\citenamefont{Prabhakar, Nepal,
  Melnik, and Kovalev}}]{PhysRevB.99.094111}
\bibinfo{author}{\bibfnamefont{S.}~\bibnamefont{Prabhakar}},
  \bibinfo{author}{\bibfnamefont{R.}~\bibnamefont{Nepal}},
  \bibinfo{author}{\bibfnamefont{R.}~\bibnamefont{Melnik}}, \bibnamefont{and}
  \bibinfo{author}{\bibfnamefont{A.~A.} \bibnamefont{Kovalev}},
  \bibinfo{journal}{Phys. Rev. B} \textbf{\bibinfo{volume}{99}},
  \bibinfo{pages}{094111} (\bibinfo{year}{2019}).

\bibitem[{\citenamefont{Brun et~al.}(2019)\citenamefont{Brun, Moreau, Somanchi,
  Nguyen, Watanabe, Taniguchi, Charlier, Stampfer, and
  Hackens}}]{PhysRevB.100.041401}
\bibinfo{author}{\bibfnamefont{B.}~\bibnamefont{Brun}},
  \bibinfo{author}{\bibfnamefont{N.}~\bibnamefont{Moreau}},
  \bibinfo{author}{\bibfnamefont{S.}~\bibnamefont{Somanchi}},
  \bibinfo{author}{\bibfnamefont{V.-H.} \bibnamefont{Nguyen}},
  \bibinfo{author}{\bibfnamefont{K.}~\bibnamefont{Watanabe}},
  \bibinfo{author}{\bibfnamefont{T.}~\bibnamefont{Taniguchi}},
  \bibinfo{author}{\bibfnamefont{J.-C.} \bibnamefont{Charlier}},
  \bibinfo{author}{\bibfnamefont{C.}~\bibnamefont{Stampfer}}, \bibnamefont{and}
  \bibinfo{author}{\bibfnamefont{B.}~\bibnamefont{Hackens}},
  \bibinfo{journal}{Phys. Rev. B} \textbf{\bibinfo{volume}{100}},
  \bibinfo{pages}{041401} (\bibinfo{year}{2019}).

\bibitem[{\citenamefont{Wehling et~al.}(2014)\citenamefont{Wehling,
  Black-Schaffer, and Balatsky}}]{Wehling2014}
\bibinfo{author}{\bibfnamefont{T.}~\bibnamefont{Wehling}},
  \bibinfo{author}{\bibfnamefont{A.}~\bibnamefont{Black-Schaffer}},
  \bibnamefont{and} \bibinfo{author}{\bibfnamefont{A.}~\bibnamefont{Balatsky}},
  \bibinfo{journal}{Advances in Physics} \textbf{\bibinfo{volume}{63}},
  \bibinfo{pages}{1} (\bibinfo{year}{2014}).

\bibitem[{\citenamefont{Wang et~al.}(2015)\citenamefont{Wang, Deng, Liu, and
  Liu}}]{WangNSR2015}
\bibinfo{author}{\bibfnamefont{J.}~\bibnamefont{Wang}},
  \bibinfo{author}{\bibfnamefont{S.}~\bibnamefont{Deng}},
  \bibinfo{author}{\bibfnamefont{Z.}~\bibnamefont{Liu}}, \bibnamefont{and}
  \bibinfo{author}{\bibfnamefont{Z.}~\bibnamefont{Liu}},
  \bibinfo{journal}{National Science Review} \textbf{\bibinfo{volume}{2}},
  \bibinfo{pages}{22} (\bibinfo{year}{2015}).

\bibitem[{\citenamefont{Xu et~al.}(2018)\citenamefont{Xu, Zou, Liu, and
  Cheng}}]{S136970211830097X}
\bibinfo{author}{\bibfnamefont{R.}~\bibnamefont{Xu}},
  \bibinfo{author}{\bibfnamefont{X.}~\bibnamefont{Zou}},
  \bibinfo{author}{\bibfnamefont{B.}~\bibnamefont{Liu}}, \bibnamefont{and}
  \bibinfo{author}{\bibfnamefont{H.-M.} \bibnamefont{Cheng}},
  \bibinfo{journal}{Materials Today} \textbf{\bibinfo{volume}{21}},
  \bibinfo{pages}{391} (\bibinfo{year}{2018}).

\bibitem[{\citenamefont{Mili\ifmmode \acute{c}\else
  \'{c}\fi{}evi\ifmmode~\acute{c}\else \'{c}\fi{}
  et~al.}(2019)\citenamefont{Mili\ifmmode \acute{c}\else
  \'{c}\fi{}evi\ifmmode~\acute{c}\else \'{c}\fi{}, Montambaux, Ozawa, Jamadi,
  Real, Sagnes, Lema\^{\i}tre, Le~Gratiet, Harouri, Bloch
  et~al.}}]{PhysRevX.9.031010}
\bibinfo{author}{\bibfnamefont{M.}~\bibnamefont{Mili\ifmmode \acute{c}\else
  \'{c}\fi{}evi\ifmmode~\acute{c}\else \'{c}\fi{}}},
  \bibinfo{author}{\bibfnamefont{G.}~\bibnamefont{Montambaux}},
  \bibinfo{author}{\bibfnamefont{T.}~\bibnamefont{Ozawa}},
  \bibinfo{author}{\bibfnamefont{O.}~\bibnamefont{Jamadi}},
  \bibinfo{author}{\bibfnamefont{B.}~\bibnamefont{Real}},
  \bibinfo{author}{\bibfnamefont{I.}~\bibnamefont{Sagnes}},
  \bibinfo{author}{\bibfnamefont{A.}~\bibnamefont{Lema\^{\i}tre}},
  \bibinfo{author}{\bibfnamefont{L.}~\bibnamefont{Le~Gratiet}},
  \bibinfo{author}{\bibfnamefont{A.}~\bibnamefont{Harouri}},
  \bibinfo{author}{\bibfnamefont{J.}~\bibnamefont{Bloch}},
  \bibnamefont{et~al.}, \bibinfo{journal}{Phys. Rev. X}
  \textbf{\bibinfo{volume}{9}}, \bibinfo{pages}{031010} (\bibinfo{year}{2019}).

\bibitem[{\citenamefont{Goerbig et~al.}(2008)\citenamefont{Goerbig, Fuchs,
  Montambaux, and Pi\'echon}}]{PhysRevB.78.045415}
\bibinfo{author}{\bibfnamefont{M.~O.} \bibnamefont{Goerbig}},
  \bibinfo{author}{\bibfnamefont{J.-N.} \bibnamefont{Fuchs}},
  \bibinfo{author}{\bibfnamefont{G.}~\bibnamefont{Montambaux}},
  \bibnamefont{and}
  \bibinfo{author}{\bibfnamefont{F.}~\bibnamefont{Pi\'echon}},
  \bibinfo{journal}{Phys. Rev. B} \textbf{\bibinfo{volume}{78}},
  \bibinfo{pages}{045415} (\bibinfo{year}{2008}).

\bibitem[{\citenamefont{Lu et~al.}(2016{\natexlab{a}})\citenamefont{Lu, Cuamba,
  Lin, Hao, Wang, Li, Zhao, and Ting}}]{PhysRevB.94.195423}
\bibinfo{author}{\bibfnamefont{H.-Y.} \bibnamefont{Lu}},
  \bibinfo{author}{\bibfnamefont{A.~S.} \bibnamefont{Cuamba}},
  \bibinfo{author}{\bibfnamefont{S.-Y.} \bibnamefont{Lin}},
  \bibinfo{author}{\bibfnamefont{L.}~\bibnamefont{Hao}},
  \bibinfo{author}{\bibfnamefont{R.}~\bibnamefont{Wang}},
  \bibinfo{author}{\bibfnamefont{H.}~\bibnamefont{Li}},
  \bibinfo{author}{\bibfnamefont{Y.}~\bibnamefont{Zhao}}, \bibnamefont{and}
  \bibinfo{author}{\bibfnamefont{C.~S.} \bibnamefont{Ting}},
  \bibinfo{journal}{Phys. Rev. B} \textbf{\bibinfo{volume}{94}},
  \bibinfo{pages}{195423} (\bibinfo{year}{2016}{\natexlab{a}}).

\bibitem[{\citenamefont{Zhou et~al.}(2014)\citenamefont{Zhou, Dong, Oganov,
  Zhu, Tian, and Wang}}]{PhysRevLett.112.085502}
\bibinfo{author}{\bibfnamefont{X.-F.} \bibnamefont{Zhou}},
  \bibinfo{author}{\bibfnamefont{X.}~\bibnamefont{Dong}},
  \bibinfo{author}{\bibfnamefont{A.~R.} \bibnamefont{Oganov}},
  \bibinfo{author}{\bibfnamefont{Q.}~\bibnamefont{Zhu}},
  \bibinfo{author}{\bibfnamefont{Y.}~\bibnamefont{Tian}}, \bibnamefont{and}
  \bibinfo{author}{\bibfnamefont{H.-T.} \bibnamefont{Wang}},
  \bibinfo{journal}{Phys. Rev. Lett.} \textbf{\bibinfo{volume}{112}},
  \bibinfo{pages}{085502} (\bibinfo{year}{2014}).

\bibitem[{\citenamefont{Lopez-Bezanilla and
  Littlewood}(2016)}]{PhysRevB.93.241405}
\bibinfo{author}{\bibfnamefont{A.}~\bibnamefont{Lopez-Bezanilla}}
  \bibnamefont{and} \bibinfo{author}{\bibfnamefont{P.~B.}
  \bibnamefont{Littlewood}}, \bibinfo{journal}{Phys. Rev. B}
  \textbf{\bibinfo{volume}{93}}, \bibinfo{pages}{241405}
  (\bibinfo{year}{2016}).

\bibitem[{\citenamefont{Zabolotskiy and Lozovik}(2016)}]{PhysRevB.94.165403}
\bibinfo{author}{\bibfnamefont{A.~D.} \bibnamefont{Zabolotskiy}}
  \bibnamefont{and} \bibinfo{author}{\bibfnamefont{Y.~E.}
  \bibnamefont{Lozovik}}, \bibinfo{journal}{Phys. Rev. B}
  \textbf{\bibinfo{volume}{94}}, \bibinfo{pages}{165403}
  (\bibinfo{year}{2016}).

\bibitem[{\citenamefont{Nakhaee et~al.}(2018)\citenamefont{Nakhaee, Ketabi, and
  Peeters}}]{PhysRevB.97.125424}
\bibinfo{author}{\bibfnamefont{M.}~\bibnamefont{Nakhaee}},
  \bibinfo{author}{\bibfnamefont{S.~A.} \bibnamefont{Ketabi}},
  \bibnamefont{and} \bibinfo{author}{\bibfnamefont{F.~M.}
  \bibnamefont{Peeters}}, \bibinfo{journal}{Phys. Rev. B}
  \textbf{\bibinfo{volume}{97}}, \bibinfo{pages}{125424}
  (\bibinfo{year}{2018}).

\bibitem[{\citenamefont{Cheng et~al.}(2017)\citenamefont{Cheng, Lang, Li, Liu,
  and Liu}}]{C7CP03736H}
\bibinfo{author}{\bibfnamefont{T.}~\bibnamefont{Cheng}},
  \bibinfo{author}{\bibfnamefont{H.}~\bibnamefont{Lang}},
  \bibinfo{author}{\bibfnamefont{Z.}~\bibnamefont{Li}},
  \bibinfo{author}{\bibfnamefont{Z.}~\bibnamefont{Liu}}, \bibnamefont{and}
  \bibinfo{author}{\bibfnamefont{Z.}~\bibnamefont{Liu}},
  \bibinfo{journal}{Phys. Chem. Chem. Phys.} \textbf{\bibinfo{volume}{19}},
  \bibinfo{pages}{23942} (\bibinfo{year}{2017}).

\bibitem[{\citenamefont{Sadhukhan and Agarwal}(2017)}]{PhysRevB.96.035410}
\bibinfo{author}{\bibfnamefont{K.}~\bibnamefont{Sadhukhan}} \bibnamefont{and}
  \bibinfo{author}{\bibfnamefont{A.}~\bibnamefont{Agarwal}},
  \bibinfo{journal}{Phys. Rev. B} \textbf{\bibinfo{volume}{96}},
  \bibinfo{pages}{035410} (\bibinfo{year}{2017}).

\bibitem[{\citenamefont{Jalali-Mola and Jafari}(2018)}]{PhysRevB.98.235430}
\bibinfo{author}{\bibfnamefont{Z.}~\bibnamefont{Jalali-Mola}} \bibnamefont{and}
  \bibinfo{author}{\bibfnamefont{S.~A.} \bibnamefont{Jafari}},
  \bibinfo{journal}{Phys. Rev. B} \textbf{\bibinfo{volume}{98}},
  \bibinfo{pages}{235430} (\bibinfo{year}{2018}).

\bibitem[{\citenamefont{Verma et~al.}(2017)\citenamefont{Verma, Mawrie, and
  Ghosh}}]{PhysRevB.96.155418}
\bibinfo{author}{\bibfnamefont{S.}~\bibnamefont{Verma}},
  \bibinfo{author}{\bibfnamefont{A.}~\bibnamefont{Mawrie}}, \bibnamefont{and}
  \bibinfo{author}{\bibfnamefont{T.~K.} \bibnamefont{Ghosh}},
  \bibinfo{journal}{Phys. Rev. B} \textbf{\bibinfo{volume}{96}},
  \bibinfo{pages}{155418} (\bibinfo{year}{2017}).

\bibitem[{\citenamefont{Islam and Jayannavar}(2017)}]{PhysRevB.96.235405}
\bibinfo{author}{\bibfnamefont{S.~F.} \bibnamefont{Islam}} \bibnamefont{and}
  \bibinfo{author}{\bibfnamefont{A.~M.} \bibnamefont{Jayannavar}},
  \bibinfo{journal}{Phys. Rev. B} \textbf{\bibinfo{volume}{96}},
  \bibinfo{pages}{235405} (\bibinfo{year}{2017}).

\bibitem[{\citenamefont{Zhang and Yang}(2018)}]{PhysRevB.97.235440}
\bibinfo{author}{\bibfnamefont{S.-H.} \bibnamefont{Zhang}} \bibnamefont{and}
  \bibinfo{author}{\bibfnamefont{W.}~\bibnamefont{Yang}},
  \bibinfo{journal}{Phys. Rev. B} \textbf{\bibinfo{volume}{97}},
  \bibinfo{pages}{235440} (\bibinfo{year}{2018}).

\bibitem[{\citenamefont{Champo and Naumis}(2019)}]{PhysRevB.99.035415}
\bibinfo{author}{\bibfnamefont{A.~E.} \bibnamefont{Champo}} \bibnamefont{and}
  \bibinfo{author}{\bibfnamefont{G.~G.} \bibnamefont{Naumis}},
  \bibinfo{journal}{Phys. Rev. B} \textbf{\bibinfo{volume}{99}},
  \bibinfo{pages}{035415} (\bibinfo{year}{2019}).

\bibitem[{\citenamefont{Paul et~al.}(2019)\citenamefont{Paul, Islam, and
  Saha}}]{PhysRevB.99.155418}
\bibinfo{author}{\bibfnamefont{G.~C.} \bibnamefont{Paul}},
  \bibinfo{author}{\bibfnamefont{S.~F.} \bibnamefont{Islam}}, \bibnamefont{and}
  \bibinfo{author}{\bibfnamefont{A.}~\bibnamefont{Saha}},
  \bibinfo{journal}{Phys. Rev. B} \textbf{\bibinfo{volume}{99}},
  \bibinfo{pages}{155418} (\bibinfo{year}{2019}).

\bibitem[{\citenamefont{Zhang et~al.}(2019)\citenamefont{Zhang, Shao, and
  Yang}}]{zhangJMMM2019}
\bibinfo{author}{\bibfnamefont{S.-H.} \bibnamefont{Zhang}},
  \bibinfo{author}{\bibfnamefont{D.-F.} \bibnamefont{Shao}}, \bibnamefont{and}
  \bibinfo{author}{\bibfnamefont{W.}~\bibnamefont{Yang}},
  \bibinfo{journal}{Journal of Magnetism and Magnetic Materials}
  \textbf{\bibinfo{volume}{491}}, \bibinfo{pages}{165631}
  (\bibinfo{year}{2019}).

\bibitem[{\citenamefont{Zhang et~al.}(2017{\natexlab{b}})\citenamefont{Zhang,
  Yang, and Chang}}]{ZhangPRB2017}
\bibinfo{author}{\bibfnamefont{S.-H.} \bibnamefont{Zhang}},
  \bibinfo{author}{\bibfnamefont{W.}~\bibnamefont{Yang}}, \bibnamefont{and}
  \bibinfo{author}{\bibfnamefont{K.}~\bibnamefont{Chang}},
  \bibinfo{journal}{Phys. Rev. B} \textbf{\bibinfo{volume}{95}},
  \bibinfo{pages}{075421} (\bibinfo{year}{2017}{\natexlab{b}}).

\bibitem[{\citenamefont{Sakurai}(1994)}]{SakuraiBook1994}
\bibinfo{author}{\bibfnamefont{J.~J.} \bibnamefont{Sakurai}},
  \emph{\bibinfo{title}{Modern Quantum Mechanics (Revised Edition)}}
  (\bibinfo{publisher}{Addison-Wesley Publishing Company, Inc.},
  \bibinfo{year}{1994}).

\bibitem[{\citenamefont{Griffiths}(1995)}]{GriffithsBook1995}
\bibinfo{author}{\bibfnamefont{D.~J.} \bibnamefont{Griffiths}},
  \emph{\bibinfo{title}{Introduction to Quantum Mechanics}}
  (\bibinfo{publisher}{Prentice Hall, Upper Saddle River, New Jersey 07458},
  \bibinfo{year}{1995}).

\bibitem[{\citenamefont{Cohen-Tannoudji
  et~al.}(2005)\citenamefont{Cohen-Tannoudji, Diu, and Laloe}}]{CohenBook2005}
\bibinfo{author}{\bibfnamefont{C.}~\bibnamefont{Cohen-Tannoudji}},
  \bibinfo{author}{\bibfnamefont{B.}~\bibnamefont{Diu}}, \bibnamefont{and}
  \bibinfo{author}{\bibfnamefont{F.}~\bibnamefont{Laloe}},
  \emph{\bibinfo{title}{Quantum Mechanics vol 2, 2nd ed}}
  (\bibinfo{publisher}{Wiley-VCH, New York}, \bibinfo{year}{2005}).

\bibitem[{\citenamefont{Zhang et~al.}(2016)\citenamefont{Zhang, Zhu, Yang, Lin,
  and Chang}}]{ZhangPRB2016}
\bibinfo{author}{\bibfnamefont{S.-H.} \bibnamefont{Zhang}},
  \bibinfo{author}{\bibfnamefont{J.-J.} \bibnamefont{Zhu}},
  \bibinfo{author}{\bibfnamefont{W.}~\bibnamefont{Yang}},
  \bibinfo{author}{\bibfnamefont{H.-Q.} \bibnamefont{Lin}}, \bibnamefont{and}
  \bibinfo{author}{\bibfnamefont{K.}~\bibnamefont{Chang}},
  \bibinfo{journal}{Phys. Rev. B} \textbf{\bibinfo{volume}{94}},
  \bibinfo{pages}{085408} (\bibinfo{year}{2016}).

\bibitem[{\citenamefont{Low and Appenzeller}(2009)}]{PhysRevB.80.155406}
\bibinfo{author}{\bibfnamefont{T.}~\bibnamefont{Low}} \bibnamefont{and}
  \bibinfo{author}{\bibfnamefont{J.}~\bibnamefont{Appenzeller}},
  \bibinfo{journal}{Phys. Rev. B} \textbf{\bibinfo{volume}{80}},
  \bibinfo{pages}{155406} (\bibinfo{year}{2009}).

\bibitem[{\citenamefont{Lu et~al.}(2016{\natexlab{b}})\citenamefont{Lu, Zhou,
  Chang, Guan, Chen, Jiang, Jiang, Wang, Yang, Feng et~al.}}]{lu2016}
\bibinfo{author}{\bibfnamefont{Y.}~\bibnamefont{Lu}},
  \bibinfo{author}{\bibfnamefont{D.}~\bibnamefont{Zhou}},
  \bibinfo{author}{\bibfnamefont{G.}~\bibnamefont{Chang}},
  \bibinfo{author}{\bibfnamefont{S.}~\bibnamefont{Guan}},
  \bibinfo{author}{\bibfnamefont{W.}~\bibnamefont{Chen}},
  \bibinfo{author}{\bibfnamefont{Y.}~\bibnamefont{Jiang}},
  \bibinfo{author}{\bibfnamefont{J.}~\bibnamefont{Jiang}},
  \bibinfo{author}{\bibfnamefont{X.-s.} \bibnamefont{Wang}},
  \bibinfo{author}{\bibfnamefont{S.~A.} \bibnamefont{Yang}},
  \bibinfo{author}{\bibfnamefont{Y.~P.} \bibnamefont{Feng}},
  \bibnamefont{et~al.}, \bibinfo{journal}{npj Computational Materials}
  \textbf{\bibinfo{volume}{2}}, \bibinfo{pages}{16011}
  (\bibinfo{year}{2016}{\natexlab{b}}).

\bibitem[{\citenamefont{Banerjee et~al.}(2009)\citenamefont{Banerjee, Singh,
  Pardo, and Pickett}}]{PhysRevLett.103.016402}
\bibinfo{author}{\bibfnamefont{S.}~\bibnamefont{Banerjee}},
  \bibinfo{author}{\bibfnamefont{R.~R.~P.} \bibnamefont{Singh}},
  \bibinfo{author}{\bibfnamefont{V.}~\bibnamefont{Pardo}}, \bibnamefont{and}
  \bibinfo{author}{\bibfnamefont{W.~E.} \bibnamefont{Pickett}},
  \bibinfo{journal}{Phys. Rev. Lett.} \textbf{\bibinfo{volume}{103}},
  \bibinfo{pages}{016402} (\bibinfo{year}{2009}).

\bibitem[{\citenamefont{Bai et~al.}(2018)\citenamefont{Bai, Zhou, Wei, Qiao,
  Liu, Liu, Jiang, and He}}]{PhysRevB.97.045413}
\bibinfo{author}{\bibfnamefont{K.-K.} \bibnamefont{Bai}},
  \bibinfo{author}{\bibfnamefont{J.-J.} \bibnamefont{Zhou}},
  \bibinfo{author}{\bibfnamefont{Y.-C.} \bibnamefont{Wei}},
  \bibinfo{author}{\bibfnamefont{J.-B.} \bibnamefont{Qiao}},
  \bibinfo{author}{\bibfnamefont{Y.-W.} \bibnamefont{Liu}},
  \bibinfo{author}{\bibfnamefont{H.-W.} \bibnamefont{Liu}},
  \bibinfo{author}{\bibfnamefont{H.}~\bibnamefont{Jiang}}, \bibnamefont{and}
  \bibinfo{author}{\bibfnamefont{L.}~\bibnamefont{He}}, \bibinfo{journal}{Phys.
  Rev. B} \textbf{\bibinfo{volume}{97}}, \bibinfo{pages}{045413}
  (\bibinfo{year}{2018}).

\bibitem[{\citenamefont{Zhou et~al.}(2019)\citenamefont{Zhou, Kerelsky, Elahi,
  Wang, Habib, Sajjad, Agnihotri, Lee, Ghosh, Ross et~al.}}]{acsnano.8b09575}
\bibinfo{author}{\bibfnamefont{X.}~\bibnamefont{Zhou}},
  \bibinfo{author}{\bibfnamefont{A.}~\bibnamefont{Kerelsky}},
  \bibinfo{author}{\bibfnamefont{M.~M.} \bibnamefont{Elahi}},
  \bibinfo{author}{\bibfnamefont{D.}~\bibnamefont{Wang}},
  \bibinfo{author}{\bibfnamefont{K.~M.~M.} \bibnamefont{Habib}},
  \bibinfo{author}{\bibfnamefont{R.~N.} \bibnamefont{Sajjad}},
  \bibinfo{author}{\bibfnamefont{P.}~\bibnamefont{Agnihotri}},
  \bibinfo{author}{\bibfnamefont{J.~U.} \bibnamefont{Lee}},
  \bibinfo{author}{\bibfnamefont{A.~W.} \bibnamefont{Ghosh}},
  \bibinfo{author}{\bibfnamefont{F.~M.} \bibnamefont{Ross}},
  \bibnamefont{et~al.}, \bibinfo{journal}{ACS Nano}
  \textbf{\bibinfo{volume}{13}}, \bibinfo{pages}{2558} (\bibinfo{year}{2019}).

\bibitem[{\citenamefont{Chaves et~al.}(2019)\citenamefont{Chaves, Jim¨¦nez,
  Santos, B?ggild, and Caridad}}]{C9NR02029B}
\bibinfo{author}{\bibfnamefont{F.~A.} \bibnamefont{Chaves}},
  \bibinfo{author}{\bibfnamefont{D.}~\bibnamefont{Jim¨¦nez}},
  \bibinfo{author}{\bibfnamefont{J.~E.} \bibnamefont{Santos}},
  \bibinfo{author}{\bibfnamefont{P.}~\bibnamefont{B?ggild}}, \bibnamefont{and}
  \bibinfo{author}{\bibfnamefont{J.~M.} \bibnamefont{Caridad}},
  \bibinfo{journal}{Nanoscale} \textbf{\bibinfo{volume}{11}},
  \bibinfo{pages}{10273} (\bibinfo{year}{2019}).

\bibitem[{\citenamefont{Mannix et~al.}(2015)\citenamefont{Mannix, Zhou, Kiraly,
  Wood, Alducin, Myers, Liu, Fisher, Santiago, Guest et~al.}}]{MannixS1513}
\bibinfo{author}{\bibfnamefont{A.~J.} \bibnamefont{Mannix}},
  \bibinfo{author}{\bibfnamefont{X.-F.} \bibnamefont{Zhou}},
  \bibinfo{author}{\bibfnamefont{B.}~\bibnamefont{Kiraly}},
  \bibinfo{author}{\bibfnamefont{J.~D.} \bibnamefont{Wood}},
  \bibinfo{author}{\bibfnamefont{D.}~\bibnamefont{Alducin}},
  \bibinfo{author}{\bibfnamefont{B.~D.} \bibnamefont{Myers}},
  \bibinfo{author}{\bibfnamefont{X.}~\bibnamefont{Liu}},
  \bibinfo{author}{\bibfnamefont{B.~L.} \bibnamefont{Fisher}},
  \bibinfo{author}{\bibfnamefont{U.}~\bibnamefont{Santiago}},
  \bibinfo{author}{\bibfnamefont{J.~R.} \bibnamefont{Guest}},
  \bibnamefont{et~al.}, \bibinfo{journal}{Science}
  \textbf{\bibinfo{volume}{350}}, \bibinfo{pages}{1513} (\bibinfo{year}{2015}).

\bibitem[{\citenamefont{Feng et~al.}(2016)\citenamefont{Feng, Zhang, Zhong, Li,
  Li, Li, Cheng, Meng, Chen, and Wu}}]{FengNC2016}
\bibinfo{author}{\bibfnamefont{B.}~\bibnamefont{Feng}},
  \bibinfo{author}{\bibfnamefont{J.}~\bibnamefont{Zhang}},
  \bibinfo{author}{\bibfnamefont{Q.}~\bibnamefont{Zhong}},
  \bibinfo{author}{\bibfnamefont{W.}~\bibnamefont{Li}},
  \bibinfo{author}{\bibfnamefont{S.}~\bibnamefont{Li}},
  \bibinfo{author}{\bibfnamefont{H.}~\bibnamefont{Li}},
  \bibinfo{author}{\bibfnamefont{P.}~\bibnamefont{Cheng}},
  \bibinfo{author}{\bibfnamefont{S.}~\bibnamefont{Meng}},
  \bibinfo{author}{\bibfnamefont{L.}~\bibnamefont{Chen}}, \bibnamefont{and}
  \bibinfo{author}{\bibfnamefont{K.}~\bibnamefont{Wu}},
  \bibinfo{journal}{Nature Chemistry} \textbf{\bibinfo{volume}{8}},
  \bibinfo{pages}{563} (\bibinfo{year}{2016}).

\bibitem[{\citenamefont{Feng et~al.}(2017)\citenamefont{Feng, Sugino, Liu,
  Zhang, Yukawa, Kawamura, Iimori, Kim, Hasegawa, Li
  et~al.}}]{PhysRevLett.118.096401}
\bibinfo{author}{\bibfnamefont{B.}~\bibnamefont{Feng}},
  \bibinfo{author}{\bibfnamefont{O.}~\bibnamefont{Sugino}},
  \bibinfo{author}{\bibfnamefont{R.-Y.} \bibnamefont{Liu}},
  \bibinfo{author}{\bibfnamefont{J.}~\bibnamefont{Zhang}},
  \bibinfo{author}{\bibfnamefont{R.}~\bibnamefont{Yukawa}},
  \bibinfo{author}{\bibfnamefont{M.}~\bibnamefont{Kawamura}},
  \bibinfo{author}{\bibfnamefont{T.}~\bibnamefont{Iimori}},
  \bibinfo{author}{\bibfnamefont{H.}~\bibnamefont{Kim}},
  \bibinfo{author}{\bibfnamefont{Y.}~\bibnamefont{Hasegawa}},
  \bibinfo{author}{\bibfnamefont{H.}~\bibnamefont{Li}}, \bibnamefont{et~al.},
  \bibinfo{journal}{Phys. Rev. Lett.} \textbf{\bibinfo{volume}{118}},
  \bibinfo{pages}{096401} (\bibinfo{year}{2017}).

\bibitem[{\citenamefont{Wang et~al.}(2019)\citenamefont{Wang, L{\"u}, Wang,
  Feng, and Zheng}}]{Wang2019}
\bibinfo{author}{\bibfnamefont{Z.-Q.} \bibnamefont{Wang}},
  \bibinfo{author}{\bibfnamefont{T.-Y.} \bibnamefont{L{\"u}}},
  \bibinfo{author}{\bibfnamefont{H.-Q.} \bibnamefont{Wang}},
  \bibinfo{author}{\bibfnamefont{Y.~P.} \bibnamefont{Feng}}, \bibnamefont{and}
  \bibinfo{author}{\bibfnamefont{J.-C.} \bibnamefont{Zheng}},
  \bibinfo{journal}{Frontiers of Physics} \textbf{\bibinfo{volume}{14}},
  \bibinfo{pages}{33403} (\bibinfo{year}{2019}).

\end{thebibliography}

\end{document}